\documentclass[a4paper,journal]{IEEEtran}
\usepackage{graphicx,ifthen,stfloats}
\usepackage{psfrag,amssymb,amsthm}
\usepackage[cmex10]{amsmath}
\usepackage{cite,color,pst-plot,stfloats,pst-sigsys}
\usepackage{pstricks-add}
\usepackage[T1]{fontenc}

\newtheorem{lem}{Lemma}
\newtheorem{cor}{Corollary}

\newtheorem{prop}{Proposition}

\theoremstyle{definition}
\newtheorem{definition}{Definition}

\newcounter{excnt}
\newenvironment{example}[2][\empty]
{
\vspace*{0.3cm}
\noindent
\ifthenelse{\equal{#1}{\empty}}{\refstepcounter{excnt}\textit{Example~\theexcnt~(#2).}}{\textit{Example~\ref{#1}~(#2).}}
}
{
\vspace*{0.3cm}
}

\title{Information Measures for Deterministic Input-Output Systems}

\author{Bernhard C. Geiger,~\IEEEmembership{Student Member, IEEE}, and Gernot Kubin,~\IEEEmembership{Member, IEEE}%
\thanks{Bernhard C. Geiger (geiger@ieee.org) and Gernot Kubin (g.kubin@ieee.org) are with the Signal Processing and Speech Communication Laboratory, Graz University of Technology}%
\thanks{Parts of this work have been presented at the 2011 IEEE Int. Sym. on Wireless Communication Systems, the 2012 Int. Z\"urich Seminar on Communications, and the 2012 IEEE Information Theory Workshop.}}

\begin{document}
\newcounter{myTempCnt}

\newcommand{\x}[1]{x[#1]}
\newcommand{\y}[1]{y[#1]}

\newcommand{\pdfy}{f_Y(y)}

\newcommand{\overbar}[1]{\mkern 1.5mu\overline{\mkern-3mu#1\mkern-0.5mu}\mkern 1.5mu}

\newcommand{\ent}[1]{H(#1)}
\newcommand{\diffent}[1]{h(#1)}
\newcommand{\derate}[1]{\overbar{h}\left(\mathbf{#1}\right)}
\newcommand{\mutinf}[1]{I(#1)}
\newcommand{\ginf}[1]{I_G(#1)}
\newcommand{\kld}[2]{D(#1||#2)}
\newcommand{\kldrate}[2]{\bar{D}(\mathbf{#1}||\mathbf{#2})}
\newcommand{\binent}[1]{H_2(#1)}
\newcommand{\binentneg}[1]{H_2^{-1}\left(#1\right)}
\newcommand{\entrate}[1]{\overbar{H}(\mathbf{#1})}
\newcommand{\mutrate}[1]{\mutinf{\mathbf{#1}}}
\newcommand{\redrate}[1]{\bar{R}(\mathbf{#1})}
\newcommand{\pinrate}[1]{\vec{I}(\mathbf{#1})}
\newcommand{\loss}[2][\empty]{\ifthenelse{\equal{#1}{\empty}}{L(#2)}{L_{#1}(#2)}}
\newcommand{\lossrate}[2][\empty]{\ifthenelse{\equal{#1}{\empty}}{\overbar{L}(\mathbf{#2})}{L_{\mathbf{#1}}(\mathbf{#2})}}
\newcommand{\gain}[1]{G(#1)}
\newcommand{\atten}[1]{A(#1)}
\newcommand{\relLoss}[2][\empty]{\ifthenelse{\equal{#1}{\empty}}{l(#2)}{l_{#1}(#2)}}
\newcommand{\relLossrate}[1]{l(\mathbf{#1})}
\newcommand{\relTrans}[1]{t(#1)}
\newcommand{\partEnt}[2]{H^{#1}(#2)}

\newcommand{\dom}[1]{\mathcal{#1}}
\newcommand{\indset}[1]{\mathbb{I}\left({#1}\right)}

\newcommand{\unif}[2]{\mathcal{U}\left(#1,#2\right)}
\newcommand{\chis}[1]{\chi^2\left(#1\right)}
\newcommand{\chir}[1]{\chi\left(#1\right)}
\newcommand{\normdist}[2]{\mathcal{N}\left(#1,#2\right)}
\newcommand{\Prob}[1]{\mathrm{Pr}(#1)}
\newcommand{\Mar}[1]{\mathrm{Mar}(#1)}
\newcommand{\Qfunc}[1]{Q\left(#1\right)}

\newcommand{\expec}[1]{\mathrm{E}\left\{#1\right\}}
\newcommand{\expecwrt}[2]{\mathrm{E}_{#1}\left\{#2\right\}}
\newcommand{\var}[1]{\mathrm{Var}\left\{#1\right\}}
\renewcommand{\det}{\mathrm{det}}
\newcommand{\cov}[1]{\mathrm{Cov}\left\{#1\right\}}
\newcommand{\sgn}[1]{\mathrm{sgn}\left(#1\right)}
\newcommand{\sinc}[1]{\mathrm{sinc}\left(#1\right)}
\newcommand{\e}[1]{\mathrm{e}^{#1}}
\newcommand{\multint}{\iint{\cdots}\int}
\newcommand{\modd}[3]{((#1))_{#2}^{#3}}
\newcommand{\quant}[1]{Q\left(#1\right)}
\newcommand{\card}[1]{\mathrm{card}(#1)}
\newcommand{\diam}[1]{\mathrm{diam}(#1)}
\newcommand{\rec}[1]{r(#1)}
\newcommand{\recmap}[1]{r_{\mathrm{MAP}}(#1)}

\newcommand{\ivec}{\mathbf{i}}
\newcommand{\hvec}{\mathbf{h}}
\newcommand{\gvec}{\mathbf{g}}
\newcommand{\avec}{\mathbf{a}}
\newcommand{\kvec}{\mathbf{k}}
\newcommand{\fvec}{\mathbf{f}}
\newcommand{\vvec}{\mathbf{v}}
\newcommand{\xvec}{\mathbf{x}}
\newcommand{\Xvec}{\mathbf{X}}
\newcommand{\Xhvec}{\hat{\mathbf{X}}}
\newcommand{\xhvec}{\hat{\mathbf{x}}}
\newcommand{\xtvec}{\tilde{\mathbf{x}}}
\newcommand{\Yvec}{\mathbf{Y}}
\newcommand{\yvec}{\mathbf{y}}
\newcommand{\Zvec}{\mathbf{Z}}
\newcommand{\Svec}{\mathbf{S}}
\newcommand{\Nvec}{\mathbf{N}}
\newcommand{\Pvec}{\mathbf{P}}
\newcommand{\muvec}{\boldsymbol{\mu}}
\newcommand{\wvec}{\mathbf{w}}
\newcommand{\Wvec}{\mathbf{W}}
\newcommand{\Hmat}{\mathbf{H}}
\newcommand{\Amat}{\mathbf{A}}
\newcommand{\Fmat}{\mathbf{F}}

\newcommand{\zerovec}{\mathbf{0}}
\newcommand{\eye}{\mathbf{I}}
\newcommand{\evec}{\mathbf{i}}

\newcommand{\zeroone}{\left[\begin{array}{c}\zerovec^T\\ \eye\end{array} \right]}
\newcommand{\zerooneT}{\left[\begin{array}{cc}\zerovec & \eye\end{array} \right]}
\newcommand{\zerooneM}{\left[\begin{array}{cc}\zerovec &\zerovec^T\\\zerovec& \eye\end{array} \right]}

\newcommand{\Cxx}{\mathbf{C}_{XX}}
\newcommand{\Cx}{\mathbf{C}_{\Xvec}}
\newcommand{\Chx}{\hat{\mathbf{C}}_{\Xvec}}
\newcommand{\Cy}{\mathbf{C}_{\Yvec}}
\newcommand{\Cz}{\mathbf{C}_{\Zvec}}
\newcommand{\Cn}{\mathbf{C}_{\mathbf{N}}}
\newcommand{\Cnt}{\underline{\mathbf{C}}_{\tilde{\mathbf{N}}}}
\newcommand{\Cntm}{\underline{\mathbf{C}}_{\tilde{\mathbf{N}}}}
\newcommand{\Cxh}{\mathbf{C}_{\hat{X}\hat{X}}}
\newcommand{\rxx}{\mathbf{r}_{XX}}
\newcommand{\Cxy}{\mathbf{C}_{XY}}
\newcommand{\Cyy}{\mathbf{C}_{YY}}
\newcommand{\Cnn}{\mathbf{C}_{NN}}
\newcommand{\Cyx}{\mathbf{C}_{YX}}
\newcommand{\Cygx}{\mathbf{C}_{Y|X}}
\newcommand{\Wmat}{\underline{\mathbf{W}}}

\newcommand{\Jac}[2]{\mathcal{J}_{#1}(#2)}

\newcommand{\NN}{{N{\times}N}}
\newcommand{\perr}{P_e}
\newcommand{\perh}{\hat{\perr}}
\newcommand{\pert}{\tilde{\perr}}

\newcommand{\vecind}[1]{#1_0^n}
\newcommand{\roots}[2]{{#1}_{#2}^{(i_{#2})}}
\newcommand{\rootx}[1]{x_{#1}^{(i)}}
\newcommand{\rootn}[2]{x_{#1}^{#2,(i)}}

\newcommand{\markkern}[1]{f_M(#1)}
\newcommand{\pole}{a_1}
\newcommand{\preim}[1]{g^{-1}[#1]}
\newcommand{\preimV}[1]{\mathbf{g}^{-1}[#1]}
\newcommand{\Xmax}{\bar{X}}
\newcommand{\Xmin}{\underbar{X}}
\newcommand{\xmax}{x_{\max}}
\newcommand{\xmin}{x_{\min}}
\newcommand{\limn}{\lim_{n\to\infty}}
\newcommand{\limk}{\lim_{k\to\infty}}
\newcommand{\limX}{\lim_{\hat{\Xvec}\to\Xvec}}
\newcommand{\limx}{\lim_{\hat{X}\to X}}
\newcommand{\limXo}{\lim_{\hat{X}_1\to X_1}}
\newcommand{\sumin}{\sum_{i=1}^n}
\newcommand{\finv}{f_\mathrm{inv}}
\newcommand{\ejtheta}{\e{\jmath\theta}}
\newcommand{\khat}{\bar{k}}
\newcommand{\modeq}[1]{g(#1)}
\newcommand{\partit}[1]{\mathcal{P}_{#1}}
\newcommand{\psd}[1]{S_{#1}(\e{\jmath \theta})}
\newcommand{\borel}[1]{\mathfrak{B}(#1)}
\newcommand{\infodim}[1]{d(#1)}

\newcommand{\delay}[2]{\psblock(#1){#2}{\footnotesize$z^{-1}$}}
\newcommand{\Quant}[2]{\psblock(#1){#2}{\footnotesize$\quant{\cdot}$}}
\newcommand{\moddev}[2]{\psblock(#1){#2}{\footnotesize$\modeq{\cdot}$}}

\newcommand{\Ymat}{\underline{\Yvec}}\newcommand{\ymat}{\underline{\yvec}}
\newcommand{\Xmat}{\underline{\Xvec}}
\renewcommand{\Cx}{\underline{\mathbf{C}}_{\Xvec}}
\renewcommand{\Chx}{\underline{\hat{\mathbf{C}}}_{\Xvec}}
\renewcommand{\Cy}{\underline{\mathbf{C}}_{\Yvec}}
\renewcommand{\gvec}{g}

\maketitle

%
%

\begin{abstract}
In this work the information loss in deterministic, memoryless systems is investigated by evaluating the conditional entropy of the input random variable given the output random variable. It is shown that for a large class of systems the information loss is finite, even if the input is continuously distributed. Based on this finiteness, the problem of perfectly reconstructing the input is addressed and Fano-type bounds between the information loss and the reconstruction error probability are derived.

For systems with infinite information loss a relative measure is defined and shown to be tightly related to R\'{e}nyi information dimension. Employing another Fano-type argument, the reconstruction error probability is bounded by the relative information loss from below.

In view of developing a system theory from an information-theoretic point-of-view, the theoretical results are illustrated by a few example systems, among them a multi-channel autocorrelation receiver. 
\end{abstract}


\section{Introduction}\label{sec:intro}
When opening a textbook on linear~\cite{Oppenheim_Discrete3} or nonlinear~\cite{Khalil_Nonlinear} input-output systems, the characterizations one typically finds -- aside from the difference or differential equation defining the system -- are almost exclusively energy-centered in nature: transfer functions, input-output stability, passivity, losslessness, and the $\mathcal{L}_2$ or energy/power gain are all defined using the amplitudes (or amplitude functions) of the involved signals, therefore essentially energetic in nature. When opening a textbook on \emph{statistical} signal processing~\cite{Manolakis_Statistical} or an engineering-oriented textbook on stochastic processes~\cite{Papoulis_Probability}, one can add correlations, power spectral densities, and how they are affected by linear and nonlinear systems (e.g., the Bussgang theorem~\cite[Thm.~9-17]{Papoulis_Probability}). By this overwhelming prevalence of energetic measures and second-order statistics, it is no surprise that many problems in system theory or signal processing are formulated in terms of energetic cost functions, e.g., the mean-squared error.

What one does not find in all these books is an information-theoretic characterization of the system at hand, despite the fact that such a characterization is strongly suggested by an elementary theorem: the data processing inequality. We know that the information content of a signal (be it a random variable or a stochastic process) cannot increase by deterministic processing, just as, loosely speaking, a passive system cannot increase the energy contained in a signal\footnote{At least by not more than a finite amount, cf.~\cite{Wyatt_Passivity}.}. 

Clearly, the information lost in a system not only depends on the system but also on the signal carrying this information; the same holds for the energy lost or gained. While there is a strong connection between energy and information for Gaussian signals (entropy and entropy rate of a Gaussian signal are related to its variance and power spectral density, respectively), for non-Gaussian signals this connection degenerates to a bound by the max-entropy property of the Gaussian distribution. Energy and information of a signal therefore can behave completely differently when fed through a system. But while we have a definition of the energy loss (namely, the inverse $\mathcal{L}_2$ gain), an analysis of the information loss in a system is still lacking.

It is the purpose of this work to close this gap and to propose the information loss -- the conditional entropy of the input given the output -- as a general system characteristic, complementing the prevailing energy-centered descriptions. The choice of this conditional entropy is partly motivated by the data processing inequality (cf.~Definition~\ref{def:loss}) and justified by a recent axiomatization of information loss~\cite{Baez_InfoLoss}. 

At present we restrict ourselves to memoryless systems\footnote{Since these systems will be described by (typically) nonlinear functions, we will use ``systems'' and ``functions'' interchangeably.} operating on (multidimensional) random variables. The reason for this is that already for this comparably simple system class a multitude of questions can be asked (e.g., what happens if we lose an infinite amount of information?), to some of which we intend to present adequate answers (e.g., by the introduction of a relative measure of information loss). This manuscript can thus be regarded as a first small step towards a system theory from an information-theoretic point-of-view; we hope that the results presented here will ley the foundation for some of the forthcoming steps, such as extensions to stochastic processes or systems with memory.

\subsection{Related Work}
To the best of the authors' knowledge, very few results about the information processing behavior of deterministic input-output systems have been published. Notable exceptions are Pippenger's analysis of the information lost in the multiplication of two integer random variables~\cite{Pippenger_MultLoss} and the work of Watanabe and Abraham concerning the \emph{rate} of information loss caused by feeding a discrete-time, finite-alphabet stationary stochastic process through a static, non-injective function~\cite{Watanabe_InfoLoss}. Moreover, in~\cite{Geiger_NLDyn1starXiv} the authors made an effort to extend the results of Watanabe and Abraham to dynamical input-output systems with finite internal memory. All these works, however, focus only on discrete random variables and stochastic processes.

Slightly larger, but still focused on discrete random variables only, is the field concerning information-theoretic cost functions: The infomax principle~\cite{Linsker_Infomax}, the information bottleneck method~\cite{Tishby_InformationBottleneck} using the Kullback-Leibler divergence as a distortion function, and system design by minimizing the error entropy (e.g.,~\cite{Principe_ITLearning}) are just a few examples of this recent trend. Additionally, Lev's approach to aggregating accounting data~\cite{Lev_Accounting}, and, although not immediately evident, the work about macroscopic descriptions of multi-agent systems~\cite{Lamarche_MacroscopicDescription} belong to that category.

A system theory for neural information processing has been proposed by Johnson\footnote{Interestingly, Johnson gave a further motivation for the present work by claiming that ``Classic information theory is silent on how to use information theoretic measures (or if they can be used) to assess actual system performance''.} in~\cite{Johnson_ITNeural}. The assumptions made there (information need not be stochastic, the same information can be represented by different signals, information can be seen as a parameter of a probability distribution, etc.) suggest the use of the Kullback-Leibler divergence as a central quantity. Although these assumptions are incompatible with ours, some similarities exist (e.g., the \emph{information transfer ratio} of a cascade in~\cite{Johnson_ITNeural} and Proposition~\ref{prop:dimTrans}).

Aside from these, the closest the literature comes to these concepts is in the field of autonomous dynamical systems or iterated maps: There, a multitude of information-theoretic characterizations are used to measure the information transfer in deterministic systems. In particular, the information flow between small and large scales, caused by the folding and stretching behavior of chaotic one-dimensional maps was analyzed in~\cite{Wiegerinck_InfoFlow}; the result they present is remarkably similar to our Proposition~\ref{prop:lossPBFdiffEnt}. Another notable example is the introduction of \emph{transfer entropy} in~\cite{Schreiber_MeasuringInfoTransfer,Kaiser_InfoTransinContProcesses} to capture the information transfer between states of different (sub-)systems. An alternative measure for the information exchanged between system components was introduced in~\cite{Liang_InfoTransfer}. Information transfer between time series and spatially distinct points in the phase space of a system are discussed, e.g., in~\cite{Vastano_InformationTransferSpatiotemporal,Matsumoto_InfoFlow}. Notably, these works can be assumed to follow the spirit of Kolmogorov and Sina\"{i}, who characterized dynamical systems exhibiting chaotic behavior with entropy~\cite{Kolmogorov_Entropy1,Kolmogorov_Entropy2,Sinai_Entropy}, cf.~\cite{Downarowicz_Entropy}. 

Recently, independent from the present authors, Baez et al.~\cite{Baez_InfoLoss} took the reverse approach and formulated axioms for the information loss induced by a measure-preserving function between finite sets, such as continuity and functoriality (cf.~Proposition~\ref{prop:cascadeLoss} in this work). They show that the difference between the entropy of the input and the entropy of the output, or, equivalently, the conditional entropy of the input given the output is the only function satisfying the axioms, thus adding justification to one of our definitions.

\subsection{Outline and Contributions}
The organization of the paper is as follows: In Section~\ref{sec:elementary} we give the definitions of absolute and relative information loss and analyze their elementary properties. Among other things, we prove a connection between relative information loss and R\'{e}nyi information dimension and the additivity of absolute information loss for a cascade. We next turn to a class of systems which has finite absolute information loss for a real-valued input in Section~\ref{sec:PBFs}. A connection to differential entropy is shown as well as numerous upper bounds on the information loss. Given this finite information loss, we present Fano-type inequalities for the probability of a reconstruction error. Section~\ref{sec:uncoutPreim} deals with systems exhibiting infinite absolute information loss, e.g., quantizers and systems reducing the dimensionality of the data, to which we apply our notion of relative information loss. Presenting a similar connection between relative information loss and the reconstruction error probability establishes a link to analog compression as investigated in~\cite{Wu_Renyi}. We apply our theoretical results to two larger systems in Section~\ref{sec:systemTheory}, a communications receiver and an accumulator. It is shown that indeed both the absolute and relative measures of information loss are necessary to fully characterize even the restricted class of systems we analyze in this work. Eventually, Section~\ref{sec:outlook} is devoted to point at open issues and lay out a roadmap for further research.


\section{Definition and Elementary Properties of Information Loss and Relative Information Loss}\label{sec:elementary}

\subsection{Notation}\label{ssec:notation}
We adopt the following notation: Random variables (RVs) are represented by upper case letters (e.g., $X$), lower case letters (e.g., $x$) are reserved for (deterministic) constants or realizations of RVs. The alphabet of an RV is indicated by a calligraphic letter (e.g., $\dom{X}$). The probability distribution of an RV $X$ is denoted by $P_X$. Similarly, $P_{XY}$ and $P_{X|Y}$ denote the joint distribution of the RVs $X$ and $Y$ and the conditional distribution of $X$ given $Y$, respectively.

If $\dom{X}$ is a proper subset of the $N$-dimensional Euclidean space $\mathbb{R}^N$ and if $P_X$ is absolutely continuous w.r.t. the $N$-dimensional Lebesgue measure $\mu^N$ (in short, $P_X\ll\mu^N$), then $P_X$ possesses a probability density function (PDF) w.r.t. the Lebesgue measure, which we will denote as $f_X$. Conversely, if the probability measure $P_X$ is concentrated on an at most countable set of points, we will write $p_X$ for its probability mass function, omitting the index whenever it is clear from the context.

We deal with functions of (real-valued) RVs: If for example $(\dom{X},\mathfrak{B}_\dom{X},P_X)$ is the (standard) probability space induced by the RV $X$ and $(\dom{Y},\mathfrak{B}_\dom{Y})$ is another (standard) measurable space, and $g{:}\ \dom{X}\to\dom{Y}$ is measurable, we can define a new RV as $Y=g(X)$. The probability distribution of $Y$ is
\begin{equation}
 \forall B\in\mathfrak{B}_\dom{Y}{:}\ P_Y(B) = P_X(\preim{B})
\end{equation}
where $\preim{B}=\{x\in\dom{X}:g(x)\in B\}$ denotes the preimage of $B$ under $g$. Abusing notation, we write $P_Y(y)$ for the probability measure of a single point instead of $P_Y(\{y\})$.

In particular, if $g$ is a quantizer which induces a partition $\partit{}$ of $\dom{X}$, we write $\hat{X}$ for the quantized RV. The uniform quantization of $X$ with hypercubes of side length $\frac{1}{2^n}$ is denoted by
\begin{equation}
 \hat{X}_n = \frac{\lfloor 2^nX\rfloor}{2^n}
\end{equation}
where the floor operation is applied element-wise if $X$ is a multi-dimensional RV. The partition induced by this uniform quantizer will be denoted as\footnote{E.g., for a one-dimensional RV, the $k$-th element of $\partit{n}$ is $\hat{\dom{X}}_k^{(n)}=[\frac{k}{2^n},\frac{k+1}{2^n})$.} $\partit{n} = \{\hat{\dom{X}}_k^{(n)}\}$. Note also that the partition $\partit{n}$ gets refined with increasing $n$ (in short, $\partit{n}\succ\partit{n+1}$).

Finally, $\ent{\cdot}$, $\diffent{\cdot}$, $\binent{\cdot}$, and $\mutinf{\cdot;\cdot}$ denote the entropy, the differential entropy, the binary entropy function, and the mutual information, respectively. Unless noted otherwise, the logarithm is taken to base two, so all entropies are measured in bits.

\subsection{Information Loss}\label{ssec:elementaryLoss}
\begin{figure}
\centering
 \begin{pspicture}[showgrid=false](-1,1)(7,5)
  \psset{style=RoundCorners}
  \pssignal(1,2){xorig}{$X$}
  \psfblock[framesize=1.5 1](4,2){c}{$g$}
  \pssignal(7,2){y}{$Y$}
  \pssignal(1,5){x}{$\hat{X}_n$}
  \psfblock[framesize=1 0.75](1,3.5){oplus}{$Q$}\pssignal(3,3.5){n}{$\partit{n}$}
  \nclist[style=Arrow]{ncline}[naput]{xorig,c,y} 
  \ncline[style=Arrow]{n}{oplus}
  \nclist[style=Arrow]{ncline}[naput]{xorig,oplus,x}
  \psset{linecolor=red}
  \ncarc[style=Dash,arcangleA=90,arcangleB=90]{xorig}{x}
  \rput*(-0.25,3.5){\scriptsize\textcolor{red}{{$\mutinf{\hat{X}_n;X}$}}}
  \ncarc[style=Dash,arcangleA=30,arcangleB=50]{x}{y}
  \rput*(4.5,4.5){\scriptsize\textcolor{red}{{$\mutinf{\hat{X}_n;Y}$}}}
\end{pspicture}
\caption{Model for computing the information loss of a memoryless input-output system $g$. $Q$ is a quantizer with partition $\partit{n}$.}
\label{fig:quantization}
\end{figure}
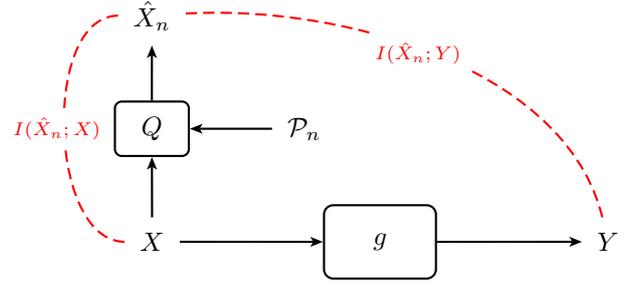

A measure of information loss in a deterministic input-output system should, roughly speaking, quantify the difference between the information available at its input and its output. While for discrete RVs this amounts to the difference of their entropies, continuous RVs require more attention. To this end, in Fig.~\ref{fig:quantization}, we propose a model to compute the information loss of a system which applies to all real-valued RVs.

In particular, we quantize the system input with partition $\partit{n}$ and compute the mutual information between the input $X$ and its quantization $\hat{X}_n$, as well as the mutual information between the system output $Y$ and $\hat{X}_n$. The first quantity is an approximation of the information available at the input, while the second approximates the information shared between input and output, i.e., the information passing through the system and thus being available at its output. By the data processing inequality (cf.~\cite[Cor.~7.16]{Gray_Entropy}), the former cannot be smaller than the latter, i.e., $\mutinf{\hat{X}_n;X}\geq\mutinf{\hat{X}_n;Y}$, with equality if the system is described by a bijective function. We compute the difference between these two mutual informations to obtain an approximation of the information lost in the system. For bijective functions, for which the two quantities are equal, the information loss will vanish, as suggested by intuition: Bijective functions describe \emph{lossless} systems.

Refining the partition yields better approximations; we thus present

\begin{definition}[Information Loss]\label{def:loss}
 Let $X$ be an RV with alphabet $\dom{X}$, and let $Y=g(X)$. The information loss induced by $g$ is
\begin{equation}
 L(X\to Y) = \limn \left(\mutinf{\hat{X}_n;X}-\mutinf{\hat{X}_n;Y}\right)=\ent{X|Y}.
\end{equation}
\end{definition}

Along the lines of~\cite[Lem.~7.20]{Gray_Entropy} one obtains
\begin{IEEEeqnarray}{RCL}
 \mutinf{\hat{X}_n;X}-\mutinf{\hat{X}_n;Y} &=& \ent{\hat{X}_n}-\ent{\hat{X}_n|X}\notag\\&&{}-\ent{\hat{X}_n}+\ent{\hat{X}_n|Y}\\&=& \ent{\hat{X}_n|Y}
\end{IEEEeqnarray}
since $\hat{X}_n$ is a function of $X$. With the monotone convergence of $\ent{\hat{X}_n|Y=y}\nearrow \ent{X|Y=y}$ implied in~\cite[Lem.~7.18]{Gray_Entropy} it follows that $\loss{X\to Y}=\ent{X|Y}$. Indeed, for a discrete input RV $X$ we obtain $\loss{X\to Y}=\ent{X}-\ent{Y}$.

While $\loss{X\to Y}$ and $\ent{X|Y}$ can be used interchangeably, we will stick to the notation $\loss{X\to Y}$ to make clear that $Y$ is a function of $X$.

For discrete input RVs $X$ or stochastic systems (e.g., communication channels) the mutual information between the input and the output $\mutinf{X;Y}$, i.e., the information transfer, is an appropriate characterization. In contrast, deterministic systems with non-discrete input RVs usually exhibit infinite information transfer $\mutinf{X;Y}$. As we will show in Section~\ref{sec:PBFs} there exists a large class of systems for which the information loss $\loss{X\to Y}$ remains finite, thus allowing to give a meaningful description of the system.

\begin{figure}[t]
 \centering
\begin{pspicture}[showgrid=false](1,1.5)(8,4)
  \psset{style=RoundCorners}
 	\pssignal(1,2){x}{$X$}
	\psfblock[framesize=1.5 1](3,2){d}{$g(\cdot)$}
	\psfblock[framesize=1.5 1](6,2){c}{$h(\cdot)$}
	\pssignal(8,2){y}{$Z$}
	\nclist[style=Arrow]{ncline}[naput]{x,d,c $Y$,y}
    \psset{style=Dash,linecolor=red,style=Arrow}
	\psline(2,2.5)(2,3)(4,3)(4,2.5)\rput[c]{0}(3,3.25){\scriptsize\textcolor{red}{$\loss{X\to Y}$}}
	\psline(5,2.5)(5,3)(7,3)(7,2.5)\rput[c]{0}(6,3.25){\scriptsize\textcolor{red}{$\loss{Y\to Z}$}}
    \psline(1.5,2.5)(1.5,3.5)(7.5,3.5)(7.5,2.5)\rput[c]{0}(4.5,3.75){\scriptsize\textcolor{red}{$\loss{X\to Z}$}}
\end{pspicture}
 \caption{Cascade of systems: The information loss of the cascade equals the sum of the individual information losses of the constituent systems.}
 \label{fig:cascade}
\end{figure}
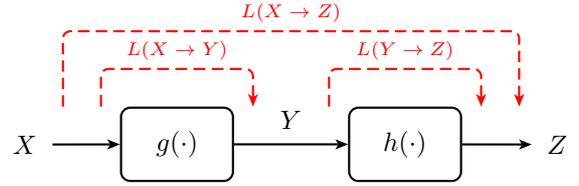

One elementary property of information loss, which will prove useful in developing a system theory from an information-theoretic point-of-view, is found in the cascade of systems (see~Fig.~\ref{fig:cascade}). We maintain\footnote{In~\cite{Baez_InfoLoss} this property was formulated as an axiom desirable for a measure of information loss.}

\begin{prop}[Information Loss of a Cascade]\label{prop:cascadeLoss}
 Consider two functions $g{:}\ \dom{X}\to\dom{Y}$ and $h{:}\ \dom{Y}\to\dom{Z}$ and a cascade of systems implementing these functions. Let $Y=g(X)$ and $Z=h(Y)$. The information loss induced by this cascade, or equivalently, by the system implementing the composition $(h\circ g)(\cdot)=h(g(\cdot))$ is given by:
\begin{equation}
\loss{X\to Z} = \loss{X\to Y} + \loss{Y\to Z}
\end{equation}
\end{prop}

\begin{IEEEproof}
 Referring to Definition~\ref{def:loss} and~\cite[Ch.~3.9]{Pinsker_InfoEngl} we obtain
\begin{IEEEeqnarray}{RCL}
 \loss{X\to Z} &=& \ent{X|Z}\\
&=& \ent{Y|Z}+\ent{X|Y,Z}\\
&=& \ent{Y|Z}+\ent{X|Y}
\end{IEEEeqnarray}
since $Y$ and $(Y,Z)$ are mutually subordinate.
\end{IEEEproof}

A sufficient condition for the information loss to be infinite is presented in
\begin{prop}[Infinite Information Loss]\label{prop:UncountInfLoss}
 Let $\gvec{:}\ \dom{X}\to\dom{Y}$, $\dom{X}\subseteq\mathbb{R}^N$, and let the input RV $X$ be such that its probability measure $P_X$ has an absolutely continuous component $P_X^{ac}\ll\mu^N$ which is supported on $\dom{X}$. If there exists a set $B\subseteq\dom{Y}$ of positive $P_Y$-measure such that the preimage $\preim{y}$ is uncountable for every $y\in B$, then
\begin{equation}
 \loss{X\to Y}=\infty.
\end{equation}
\end{prop}

\begin{IEEEproof}
We write the information loss $\loss{X\to Y}$ as
\begin{IEEEeqnarray}{RCL}
\loss{X\to Y} &=& \ent{X|Y} = \int_\dom{Y} \ent{X|Y=y}dP_Y(y)\\
&\geq& \int_B \ent{X|Y=y}dP_Y(y)
\end{IEEEeqnarray}
since $B\subseteq\dom{Y}$. By assumption, the conditional probability measure $P_{X|Y=y}$ is not concentrated on a countable set of points (the preimage of $y$ under $g$ is uncountable, and the probability measure $P_X$ has an absolutely continuous component on all $\dom{X}$) one obtains $\ent{X|Y=y}=\infty$ for all $y\in B$. The proof follows from $P_Y(B)>0$.
\end{IEEEproof}

It will be useful to explicitly state the following

\begin{cor}\label{cor:UncountInfLoss}
 Let $P_X\ll\mu^N$. If there exists a point $y^*\in\dom{Y}$ such that $P_Y(y^*)>0$, then
\begin{equation}
 \loss{X\to Y}=\infty.
\end{equation}
\end{cor}

\begin{IEEEproof}
Since $y^*$ has positive $P_Y$-measure and since $P_X\ll\mu^N$, the preimage of $y^*$ under $g$ needs to be uncountable.
\end{IEEEproof}

In other words, if the input is continuously distributed and if the distribution of the output has a non-vanishing discrete component, the information loss is infinite. A particularly simple example for such a case is a quantizer:

\begin{example}{Quantizer}\label{ex:quantizer}
 We now look at the information loss of a scalar quantizer, i.e., of a system described by a function
\begin{equation}
 g(x) = \lfloor x\rfloor.
\end{equation}
With the notation introduced above we obtain $Y=\hat{X}_0=g(X)$. Assuming that $X$ has an absolutely continuous distribution ($P_X\ll\mu$), there will be at least one point $y^*$ for which $\Prob{Y=y^*}=P_Y(y^*)=P_X([y^*,y^*+1))>0$. The conditions of Corollary~\ref{cor:UncountInfLoss} are thus fulfilled and we obtain
\begin{equation}
 \loss{X\to\hat{X}_0}=\infty.
\end{equation}
\end{example}

This simple example illustrates the information loss as the difference between the information available at the input and the output of a system: While in all practically relevant cases a quantizer will always have finite information at its output ($\ent{Y}<\infty$), the information at the input is infinite as soon as $P_X$ has a continuous component. 

While for a quantizer the mutual information between the input and the output may be a more appropriate characterization because it remains finite, the following example shows that also mutual information has its limitations:

\begin{figure}[t!]
  \begin{center}
 \begin{pspicture}[showgrid=false](-2,-2)(2,2)
	\psaxeslabels{->}(0,0)(-2,-2)(2,2){$x$}{$g(x)$}
  \psplot[style=Graph,linecolor=black,plotpoints=500]{-1.8}{-0.85}{x}
	\psplot[style=Graph,linecolor=black,plotpoints=500]{0.85}{1.8}{x}
	\psplot[style=Graph,linecolor=black,plotpoints=500]{-0.8}{.8}{0}
	\psdisk[fillcolor=black](0.8,0){0.07}\psdisk[fillcolor=black](-0.8,0){0.07}
	\pscircle(0.8,0.8){0.07}\pscircle(-0.8,-0.8){0.07}
	\psTick{90}(0.8,0) \rput(0.8,-0.3){$c$}
	\psTick{90}(-0.8,0) \rput(-0.8,-0.3){$-c$}
 \end{pspicture}
\end{center}
\caption{The center clipper -- an example for a system with both infinite information loss and information transfer.}
\label{fig:cclipper}
\end{figure}
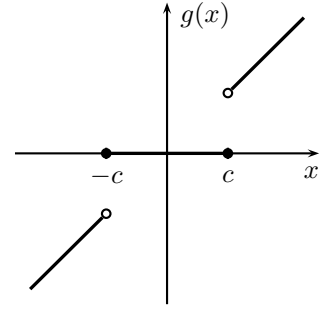

\begin{example}{Center Clipper}\label{ex:cclipper}
The center clipper, used for, e.g., residual echo suppression~\cite{Vary_DigSpeechTransm}, can be described by the following function (see Fig.~\ref{fig:cclipper}):
\begin{equation}
 g(x) = \begin{cases}
 x,& \text{ if }|x|>c\\ 0, & \text{ otherwise}
        \end{cases}
\end{equation}
Assuming again that $P_X\ll\mu$ and that $0<P_X([-c,c])<1$, with Corollary~\ref{cor:UncountInfLoss} the information loss becomes infinite. On the other hand, there exists a subset of $\dom{X}\times\dom{Y}$ (namely, a subset of the line $x=y$) with positive $P_{XY}$ measure, for which $P_XP_Y$ vanishes. Thus, with~\cite[Thm.~2.1.2]{Pinsker_InfoEngl}
\begin{equation}
 \mutinf{X;Y}=\infty.
\end{equation}
\end{example}

\subsection{Relative Information Loss}\label{ssec:elementaryRelLoss}
As the previous example shows, there are systems for which neither information transfer (i.e., the mutual information between input and output) nor information loss provides sufficient insight. For these systems, a different characterization is necessary, which leads to

\begin{definition}[Relative Information Loss]\label{def:relloss}
 The \emph{relative information loss} induced by $Y=g(X)$ is defined as
\begin{equation}
 \relLoss{X\to Y} = \limn \frac{\ent{\hat{X}_n|Y}}{\ent{\hat{X}_n}}
\end{equation}
provided the limit exists.
\end{definition}

One elementary property of the relative information loss is that \mbox{$\relLoss{X\to Y}\in[0,1]$}, due to the non-negativity of entropy and the fact that $\ent{\hat{X}_n|Y}\leq \ent{\hat{X}_n}$. The relative information loss is related to the R\'{e}nyi information dimension, which we will establish after presenting

\begin{definition}[R\'{e}nyi Information Dimension~\cite{Renyi_InfoDim}]\label{def:infodim}
The \emph{information dimension} of an RV $X$ is
\begin{equation}
 \infodim{X} = \limn \frac{\ent{\hat{X}_n}}{n}
\end{equation}
provided the limit exists and is finite.
\end{definition}

We adopted this definition from Wu and Verd\'{u}, who showed in~\cite[Prop.~2]{Wu_Renyi} that it is equivalent to the one given by R\'{e}nyi in~\cite{Renyi_InfoDim}. Note further that we excluded the case that the information dimension is infinite, which may occur if $\ent{\hat{X}_0}=\infty$~\cite[Prop.~1]{Wu_Renyi}. Conversely, if the information dimension of an RV $X$ exists, it is guaranteed to be finite if $\ent{\hat{X}_0}<\infty$~\cite{Renyi_InfoDim} or if $\expec{|X|^\epsilon}<\infty$ for some $\epsilon>0$~\cite{Wu_Renyi}. Aside from that, the information dimension exists for discrete RVs and RVs with probability measures absolutely continuous w.r.t. the Lebesgue measure on a sufficiently smooth manifold~\cite{Renyi_InfoDim}, for mixtures of RVs with existing information dimension~\cite{Renyi_InfoDim,Wu_Renyi,Smieja_EntropyMixture}, and self-similar distributions generated by iterated function systems~\cite{Wu_Renyi}. Finally, the information dimension exists if the MMSE dimension exists~\cite[Thm.~8]{Verdu_MMSE}. For the remainder of this work we will assume that the information dimension of all considered RVs exists and is finite. 

We are now ready to state
\begin{prop}[Relative Information Loss and Information Dimension]\label{prop:RILDim}
 Let $X$ be an $N$-dimensional RV with positive information dimension $\infodim{X}$. If $\infodim{X|Y=y}$ exists and is finite $P_Y$-a.s., the relative information loss equals
\begin{equation}
 \relLoss{X \to Y} = \frac{\infodim{X|Y}}{\infodim{X}}
\end{equation}
where $\infodim{X|Y}=\int_{\dom{Y}}\infodim{X|Y=y}dP_Y(y)$.
\end{prop}

\begin{IEEEproof}
From Definition~\ref{def:relloss} we obtain
\begin{IEEEeqnarray}{RCL}
\relLoss{X\to Y} 
&=& \limn \frac{\int_\dom{Y} \ent{\hat{X}_n|Y=y}dP_Y(y) }{\ent{\hat{X}_n}}\\
&=& \limn \frac{\int_\dom{Y} \frac{\ent{\hat{X}_n|Y=y}}{ n}dP_Y(y)  }{\frac{\ent{\hat{X}_n}}{ n}}.
\end{IEEEeqnarray}
By assumption, the limit of the denominator and the expression under the integral both exist and correspond to $\infodim{X}$ and $\infodim{X|Y=y}$, respectively. Since for an $\mathbb{R}^N$-valued RV $X$ the information dimension satisfies~\cite{Wu_Renyi,Renyi_InfoDim}
\begin{equation}
 \infodim{X|Y=y} \leq N
\end{equation}
one can apply Lebesgue's dominated convergence theorem (e.g.,~\cite{Rudin_Analysis3}) to exchange the order of the limit and the integral. The limit of the numerator thus exists and we continue with
\begin{multline}
 \relLoss{X \to Y} = \frac{\limn\int_\dom{Y} \frac{\ent{\hat{X}_n|Y=y}}{ n}dP_Y(y)  }{\infodim{X}}\\
  =\frac{\int_\dom{Y} \infodim{X|Y=y}dP_Y(y)  }{\infodim{X}}= \frac{\infodim{X|Y}}{\infodim{X}}.
\end{multline}
This completes the proof.
\end{IEEEproof}

We accompany the relative information loss by its logical complement, the relative information transfer:
\begin{definition}[Relative Information Transfer]\label{def:relTrans}
  The \emph{relative information transfer} achieved by $Y=g(X)$ is
\begin{equation}
 \relTrans{X\to Y} = \limn \frac{\mutinf{\hat{X}_n;Y}}{\ent{\hat{X}_n}} = 1 -\relLoss{X\to Y}
\end{equation}
provided the limit exists.
\end{definition}

While, as suggested by the data processing inequality, the focus of this work is on information loss, we introduce this definition to simplify a few proofs of the forthcoming results. In particular, we maintain

\begin{prop}[Relative Information Transfer and Information Dimension]\label{prop:dimTrans}
 Let $X$ be an RV with positive information dimension $\infodim{X}$ and let $g$ be a Lipschitz function. Then, the relative information transfer through this function is
\begin{equation}
 \relTrans{X\to Y} = \frac{\infodim{Y}}{\infodim{X}}.
\end{equation}
\end{prop}

\begin{IEEEproof}
 See Appendix~\ref{app:proofLip}.
\end{IEEEproof}

This result for Lipschitz functions is the basis for several of the elementary properties of relative information loss presented in the remainder of this section. Aside from that, it suggests that, at least for this restricted class of functions,
\begin{equation}
 \infodim{X} = \infodim{X,Y} = \infodim{X|Y}+\infodim{Y}
\end{equation}
holds. This complements the results of~\cite{Cutler_ChainRule}, where it was shown that the \emph{point-wise} information dimension satisfies this chain rule (second equality) given that the conditional probability measure satisfies a Lipschitz property. Moreover, the first equality was shown to hold for the point-wise information dimension given that $Y$ is a Lipschitz function of $X$; for non-Lipschitz functions the point-wise information dimension of $(X,Y)$ may exceed the one of $X$. If the same holds for the information dimension (which is the expectation over the point-wise information dimension) is an interesting question for future research.

We can now present the counterpart of Proposition~\ref{prop:cascadeLoss} for relative information loss:
\begin{prop}[Relative Information Loss of a Cascade]\label{prop:cascadeRelLoss}
Consider two Lipschitz functions $\gvec{:}\ \dom{X}\to\dom{Y}$ and $h{:}\ \dom{Y}\to\dom{Z}$ and a cascade of systems implementing these functions. Let $Y=g(X)$ and $Z=h(Y)$. For the cascade of these systems the relative information transfer and relative information loss are given as
\begin{equation}
 \relTrans{X\to Z}=\relTrans{X\to Y}\relTrans{Y\to Z}
\end{equation}
and
\begin{multline}
 \relLoss{X\to Z}\\ = \relLoss{X\to Y}+\relLoss{Y\to Z} - \relLoss{X\to Y}\relLoss{Y\to Z}
\end{multline}
respectively.
\end{prop}

\begin{IEEEproof}
 The proof follows from Proposition~\ref{prop:dimTrans} and Definition~\ref{def:relTrans}.
\end{IEEEproof}

\subsection{Interplay between Information Loss and Relative Information Loss}\label{ssec:interplay}
We introduced the relative information loss to characterize systems for which the absolute information loss from Definition~\ref{def:loss} is infinite. The following result shows that, at least for input RVs with infinite entropy, an infinite absolute information loss is a prerequisite for positive relative information loss:

\begin{prop}[Positive Relative Loss leads to Infinite Absolute Loss]\label{prop:RelInfLoss}
 Let $X$ be such that $\ent{X}=\infty$ and let\\ $\relLoss{X\to Y}>0$. Then, $\loss{X\to Y}=\infty$.
\end{prop}

\begin{IEEEproof}
 We prove the proposition by contradiction. To this end, assume that $\loss{X\to Y}=\ent{X|Y}=\kappa<\infty$. Thus,
\begin{IEEEeqnarray}{RCL}
 \relLoss{X \to Y} &=& \limn \frac{\ent{\hat{X}_n|Y}}{\ent{\hat{X}_n}}\\
&\stackrel{(a)}{\leq}& \limn \frac{\ent{X|Y}}{\ent{\hat{X}_n}}\\
&\stackrel{(b)}{=} & 0
\end{IEEEeqnarray}
where $(a)$ is due to data processing and $(b)$ follows from $\ent{X|Y}=\kappa<\infty$ and from $\ent{\hat{X}_n}\to\ent{X}=\infty$ (e.g.,~\cite[Lem.~7.18]{Gray_Entropy}).
\end{IEEEproof}

Note that the converse is not true: There exist examples where an infinite amount of information is lost, but for which the relative information loss nevertheless vanishes, i.e., $\relLoss{X\to Y}=0$ (see Example~\ref{ex:infloss} further below).

\section{Information Loss for Piecewise Bijective Functions}\label{sec:PBFs}
In this section we analyze the information loss for a restricted class of functions and under the practically relevant assumption that the input RV has a probability distribution $P_X\ll\mu^N$ supported on $\dom{X}$. Let $\{\dom{X}_i\}$ be a partition of $\dom{X}\subseteq\mathbb{R}^N$, i.e., the elements $\dom{X}_i$ are disjoint and unite to $\dom{X}$, and let $P_X(\dom{X}_i)>0$ for all $i$. We present

\begin{definition}[Piecewise Bijective Function]\label{def:PBF}
A piecewise bijective function $\gvec{:}\  \dom{X}\to\dom{Y}$, $\dom{X},\dom{Y}\subseteq\mathbb{R}^N$, is a surjective function defined in a piecewise manner:
\begin{equation}
 g(x) = \begin{cases}
  g_1(x), & \text{if } x \in\dom{X}_1\\
  g_2(x), & \text{if } x \in\dom{X}_2\\
  \vdots
 \end{cases}
\end{equation}
where each $\gvec_i{:}\ \dom{X}_i\to\dom{Y}_i$ is bijective. Furthermore, the Jacobian matrix $\Jac{g}{\cdot}$ exists on the closures of $\dom{X}_i$, and its determinant, $\det\Jac{g}{\cdot}$, is non-zero $P_X$-a.s.
\end{definition}

A direct consequence of this definition is that also $P_Y\ll\mu^N$. Thus, $P_Y$ possesses a PDF $f_Y$ w.r.t. the Lebesgue measure $\mu^N$ which, using the method of transformation (e.g.~\cite[p.~244]{Papoulis_Probability}, can be computed as
\begin{equation}
 f_Y(y) = \sum_{x_i\in\preim{y}} \frac{f_X(x_i)}{|\det\Jac{g}{x_i}|}.
\end{equation}

In addition to that, since the preimage $\preim{y}$ is countable for all $y\in\dom{Y}$, it follows that $\infodim{X|Y}=0$. But with $\infodim{X}=N$ from $P_X\ll\mu^N$ we can apply Proposition~\ref{prop:RILDim} to obtain $\relLoss{X\to Y}=0$. Thus, relative information loss will not tell us much about the behavior of the system. In the following, we therefore stick to Definition~\ref{def:loss} and analyze the (absolute) information loss in piecewise bijective functions (PBFs).

\subsection{Information Loss in PBFs}\label{ssec:lossPBFs}
We present
\begin{prop}[Information Loss and Differential Entropy]\label{prop:lossPBFdiffEnt}
 The information loss induced by a PBF is given as
\begin{equation}
 \loss{X\to Y} = \diffent{X}-\diffent{Y} + \expec{\log|\det\Jac{g}{X}|}
\end{equation}
where the expectation is taken w.r.t. $X$.
\end{prop}

\def \proofILDIFFhere {0}
\begin{IEEEproof}
\ifthenelse{\proofILDIFFhere=1}{Recall that Definition~\ref{def:loss} states
\begin{equation}
 \loss{X\to Y} = \limn \left(\mutinf{\hat{X}_n;X}-\mutinf{\hat{X}_n;Y}\right).
\end{equation}
Let $\hat{X}_n=\hat{x}_k$ if $x\in\hat{\dom{X}}_k^{(n)}$.
The conditional probability measure $P_{X|\hat{X}_n=\hat{x}_k}\ll\mu^N$ and thus possesses a density
\begin{equation}
 f_{X|\hat{X}_n}(x,\hat{x}_k) = \begin{cases}
                                 \frac{f_X(x)}{p(\hat{x}_k)},& \text{ if } x\in\hat{\dom{X}}_k^{(n)}\\0,&\text{ else}
                                \end{cases}
\end{equation}
where $p(\hat{x}_k)=P_X(\hat{\dom{X}}_k^{(n)})>0$. By the same arguments as in the beginning of Section~\ref{sec:PBFs}, also $P_{Y|\hat{X}_n=\hat{x}_k}\ll\mu^N$, and its PDF if given by the method of transformation.

With~\cite[Ch.~8.5]{Cover_Information2},
\begin{multline}
 \loss{X\to Y}\\ = \limn \left(\diffent{X}-\diffent{X|\hat{X}_n}-\diffent{Y}+\diffent{Y|\hat{X}_n}\right)\\
= \diffent{X}-\diffent{Y}+\limn\left(\diffent{Y|\hat{X}_n}-\diffent{X|\hat{X}_n}\right)
\end{multline}
The latter difference can be written as
\begin{multline}
 \diffent{Y|\hat{X}_n}-\diffent{X|\hat{X}_n}\\ = \sum_{\hat{x}_k} p(\hat{x}_k) \left(\diffent{Y|\hat{X}_n=\hat{x}_k}-\diffent{X|\hat{X}_n=\hat{x}_k}\right).
\end{multline}
Since (see~\cite[Thm.~5-1]{Papoulis_Probability})
\begin{multline}
 \diffent{Y|\hat{X}_n=\hat{x}_k} = -\int_\dom{Y} f_{Y|\hat{X}_n}(y,\hat{x}_k)\log f_{Y|\hat{X}_n}(y,\hat{x}_k) dy\\
=-\int_\dom{X} f_{X|\hat{X}_n}(x,\hat{x}_k)\log f_{Y|\hat{X}_n}(g(x),\hat{x}_k) dx\\
=-\frac{1}{p(\hat{x}_k)}\int_{\hat{\dom{X}}_k^{(n)}} f_{X}(x)\log f_{Y|\hat{X}_n}(g(x),\hat{x}_k) dx
\end{multline}
we obtain
\begin{multline}
 \diffent{Y|\hat{X}_n}-\diffent{X|\hat{X}_n}\\ = \sum_{k}\int_{\hat{\dom{X}}_k^{(n)}} f_{X}(x)\log \frac{f_{X|\hat{X}_n}(x,\hat{x}_k)}{f_{Y|\hat{X}_n}(g(x),\hat{x}_k)} dx.\label{eq:diffdiffents}
\end{multline}
Finally, using the method of transformation,
\begin{equation}
 f_{Y|\hat{X}_n}(g(x),\hat{x}_k) = \sum_{x_i\in\preim{g(x)}} \frac{f_{X|\hat{X}_n}(x_i,\hat{x}_k)}{|\det\Jac{g}{x_i}|}.\label{eq:Ycond}
\end{equation}

Since the preimage of $g(x)$ is a set separated by neighborhoods\footnote{The space $\mathbb{R}^N$ is Hausdorff, so any two distinct points are separated by neighborhoods.}, we can find an $n_0$ such that 
\begin{equation}
\forall n\geq n_0{:}\ k=\{\hat{k}: x \in \hat{\dom{X}}_{\hat{k}}^{(n)}\}{:}\   \preim{g(x)} \cap \hat{\dom{X}}_k^{(n)} = x
\end{equation}
i.e., such that from this index on, the element of the partition under consideration, $\hat{\dom{X}}_k^{(n)}$, contains just a single element of the preimage, $x$. Since $f_{X|\hat{X}_n}$ is non-zero only for arguments in $\hat{\dom{X}}_k^{(n)}$, in this case~\eqref{eq:Ycond} degenerates to
\begin{equation}
 f_{Y|\hat{X}_n}(g(x),\hat{x}_k) = \frac{f_{X|\hat{X}_n}(x,\hat{x}_k)}{|\det\Jac{g}{x}|}.
\end{equation}
Consequently, the ratio
\begin{equation}
 \frac{f_{X|\hat{X}_n}(x,\hat{x}_k)}{f_{Y|\hat{X}_n}(g(x),\hat{x}_k)}\nearrow |\det\Jac{g}{x}|
\end{equation}
monotonically (the number of positive terms in the sum in the denominator reduces with $n$). We can thus apply the monotone convergence theorem, e.g.,~\cite[pp.~21]{Rudin_Analysis3} and obtain
\begin{multline}
 \limn \diffent{Y|\hat{X}_n}-\diffent{X|\hat{X}_n}\\ = \sum_{k}\int_{\hat{\dom{X}}_k^{(n)}} f_{X}(x)\log|\det\Jac{g}{x}| dx\\
=\expec{\log|\det\Jac{g}{X}|}.
\end{multline}
This completes the proof.\endproof}{See Appendix~\ref{app:proofILDIFF}.}
\end{IEEEproof}

Aside from being one of the main results of this work, it also complements a result presented in~\cite[pp.~660]{Papoulis_Probability}. There, it was claimed that
\begin{equation}
 \diffent{Y} \leq \diffent{X} + \expec{\log|\det\Jac{g}{X}|}\label{eq:papDiff}
\end{equation}
where equality holds if and only if $\gvec$ is bijective, i.e., a lossless system. This inequality results from
\begin{equation}
 f_Y(g(x))\geq \frac{f_X(x)}{|\det\Jac{g}{x}|}
\end{equation}
with equality if and only if $g$ is invertible at $x$. Proposition~\ref{prop:lossPBFdiffEnt} essentially states that the difference between the right-hand side and the left-hand side of~\eqref{eq:papDiff} is the information lost due to data processing.

\begin{example}{Square-Law Device and Gaussian Input}\label{ex:square}
 We illustrate this result by assuming that $X$ is a zero-mean, unit variance Gaussian RV and that $Y=X^2$. We switch in this example to measuring entropy in nats, so that we can compute the differential entropy of $X$ as $\diffent{X}=\frac{1}{2}\ln(2\pi\e{})$. The output $Y$ is a $\chi^2$-distributed RV with one degree of freedom, for which the differential entropy can be computed as~\cite{Lazo_Entropies}
\begin{equation}
 \diffent{Y} = \frac{1}{2}\left(1+\ln \pi-\gamma\right)
\end{equation}
where $\gamma$ is the Euler-Mascheroni constant~\cite[pp.~3]{Abramowitz_Handbook}. The Jacobian determinant degenerates to the derivative, and using some calculus we obtain
\begin{equation}
 \expec{\ln|g'(X)|} = \expec{\ln |2X|} = \frac{1}{2}\left(\ln 2-\gamma\right).
\end{equation}
Applying Proposition~\ref{prop:lossPBFdiffEnt} we obtain an information loss of $\loss{X\to Y}=\ln 2$, which after changing the base of the logarithm amounts to one bit. Indeed, the information loss induced by a square-law device is always one bit if the PDF of the input RV has even symmetry~\cite{Geiger_ISIT2011arXiv}.
\end{example}

We note in passing that the statement of Proposition~\ref{prop:lossPBFdiffEnt} has a tight connection to the theory of iterated function systems. In particular,~\cite{Wiegerinck_InfoFlow} analyzed the \emph{information flow} in one-dimensional maps, which is the difference between information generation via stretching (corresponding to the term involving the Jacobian determinant) and information reduction via folding (corresponding to information loss). Ruelle~\cite{Ruelle_EntropyProduction} later proved that for a restricted class of systems the \emph{folding entropy} ($\loss{X\to Y}$) cannot fall below the information generated via stretching, and therefore speaks of \emph{positivity of entropy production}. He also established a connection to the Kolmogorov-Sina\"{i} entropy rate. In~\cite{Jost_DynSys} both components constituting information flow in iterated function systems are described as ways a dynamical system can lose information. Since the connection between the theory of iterated function maps, Kolmogorov-Sina\"{i} entropy rate, and information loss deserves undivided attention, we leave a more thorough analysis thereof for a later time.

We turn to an explanation of why the information loss actually occurs. Intuitively, the information loss is due to the non-injectivity of $g$, i.e., employing Definition~\ref{def:loss}, due to the fact that the bijectivity of $g$ is only piecewise. We will make this precise after introducing
\begin{definition}[Partition Indicator]\label{def:W}
 Let $W$ be a discrete RV which is defined as
\begin{equation}
 W=i \ \ \text{if} \ \ X\in\dom{X}_i
\end{equation}
for all $i$.
\end{definition}

\begin{prop}\label{prop:partitionLoss}
 The information loss is identical to the uncertainty about the set $\dom{X}_i$ from which the input was taken, i.e.,
\begin{equation}
 \loss{X\to Y} = \ent{W|Y}.
\end{equation}
\end{prop}

\def \proofWhere {0}
\begin{IEEEproof}
\ifthenelse{\proofWhere=1}{ We start by writing
\begin{multline}
 \ent{W|Y} = \int_{\dom{Y}} \ent{W|Y=y}dP_Y(y)\\
= -\int_\dom{Y} \sum_i p(i|y)\log p(i|y) f_Y(y) dy\label{eq:condW}
\end{multline}
where
\begin{equation}
 p(i|y) = \Prob{W=i|Y=y} = P_{X|Y=y}(\dom{X}_i).
\end{equation}
We now, for the sake of simplicity, permit the Dirac delta distribution $\delta$ as a PDF for discrete (atomic) probability measures. In particular and following~\cite{Chi_TransformingDirac}, we write for the conditional PDF of $Y$ given $X=x$,
\begin{equation}
 f_{Y|X}(x,y) = \delta(y-g(x)) = \sum_{x_i\in\preim{y}}\frac{\delta(x-x_i)}{|\det\Jac{g}{x_i}|}.
\end{equation}
Applying Bayes' theorem for densities we get
\begin{align}
 p(i|y) &= \int_{\dom{X}_i} f_{X|Y}(x,y) dx\\
&=\int_{\dom{X}_i} \frac{f_{Y|X}(x,y)f_X(x)}{f_Y(y)} dx\\
&= \frac{1}{f_Y(y)} \int_{\dom{X}_i} \sum_{x_k\in\preim{y}}\frac{\delta(x-x_k)}{|\det\Jac{g}{x_k}|} f_X(x) dx\\
&= \begin{cases}
     \frac{f_X(g_i^{-1}(y))}{|\det\Jac{g}{g_i^{-1}(y)}|f_Y(y)}, & \text{ if } y\in\dom{Y}_i\\
0,&\text{ if } y\notin\dom{Y}_i
    \end{cases}
\end{align}
by the properties of the delta distribution (e.g.,~\cite{Papoulis_Fourier}) and since, by Definition~\ref{def:PBF}, at most one element of the preimage of $y$ lies in $\dom{X}_i$.

We rewrite~\eqref{eq:condW} as
\begin{IEEEeqnarray}{RCL}
 \ent{W|Y}  &=& -\sum_i\int_{\dom{Y}_i}  p(i|y)\log p(i|y) f_Y(y) dy
\end{IEEEeqnarray}
after exchanging the order of summation and integration with the help of Tonelli's theorem~\cite[Thm.~2.37]{Folland_Analysis} and by noticing that $p(i|y)=0$ if $y\notin\dom{Y}_i$. Inserting the expression for $p(i|y)$ and changing the integration variables by substituting $x=g_i^{-1}(y)$ in each integral yields
\begin{align}
 &\ent{W|Y}\notag \\&= -\sum_i\int_{\dom{Y}_i}  \frac{f_X(g_i^{-1}(y))}{|\det\Jac{g}{g_i^{-1}(y)}|}\log  \frac{f_X(g_i^{-1}(y))}{|\det\Jac{g}{g_i^{-1}(y)}|f_Y(y)} dy\\
&=-\sum_i\int_{\dom{X}_i}  f_X(x)\log  \frac{f_X(x)}{|\det\Jac{g}{x}|f_Y(g(x))} dx\\
&=-\int_{\dom{X}}  f_X(x)\log  \frac{f_X(x)}{|\det\Jac{g}{x}|f_Y(g(x))} dx\\
&\stackrel{(a)}{=} \diffent{X}-\diffent{Y}+\expec{\log|\det\Jac{g}{X}|}\\
&= \loss{X\to Y}
\end{align}
where $(a)$ is due to splitting the logarithm and applying~\cite[Thm.~5-1]{Papoulis_Probability}.\endproof}{See Appendix~\ref{app:proofW}.}
\end{IEEEproof}

This proposition states that the information loss of a PBF stems from the fact that by observing the output one has remaining uncertainty about which element of the partition $\{\dom{X}_i\}$ contained the input value. Moreover, by looking at Definition~\ref{def:W} one can see that $W$ is obtained by quantizing $X$ with partition $\{\dom{X}_i\}$. Consequently, the limit in Definition~\ref{def:loss} is actually achieved at a comparably coarse partition.

The result permits a simple, but interesting
\begin{cor}\label{cor:partitionLoss}
 $Y$ and $W$ together determine $X$, i.e.,
\begin{equation}
 \ent{X|Y,W}=0.
\end{equation}
\end{cor}

\begin{IEEEproof}
Since $W$ is obviously a function of $X$,
\begin{multline}
 \loss{X\to Y} = \ent{X|Y}=\ent{X,W|Y}\\
=\ent{X|W,Y}+\ent{W|Y}\\
=\ent{X|W,Y}+\loss{X\to Y}
\end{multline}
 from which $\ent{X|Y,W}=0$ follows.
\end{IEEEproof}
In other words, knowing the output value, and the element of the partition from which the input originated, perfect reconstruction is possible. We will make use of this in Section~\ref{ssec:reconLoss}.

Before proceeding, we present an example where an infinite amount of information is lost in a PBF:

\begin{example}{Infinite Loss}\label{ex:infloss}
\begin{figure}[t]
  \centering
    \begin{pspicture}[showgrid=false](-4,-3.5)(4,3)
      \psframe[framearc=0,fillstyle=solid,fillcolor=red!20,linecolor=red!20](-3,0)(-2.813,3)
      \psTick{90}(3,0) \rput[th](3,-0.1){\footnotesize$1$}
      \psTick{90}(0,0) \rput[th](0,-0.1){\footnotesize$\frac{1}{2}$}
      \psTick{90}(-1.5,0) \rput[th](-1.5,-0.1){\footnotesize$\frac{1}{4}$}
      \psTick{90}(-2.25,0) \rput[th](-2.25,-0.1){\footnotesize$\frac{1}{8}$}
      \psTick{90}(-2.625,0) \rput[th](-2.625,-0.1){\footnotesize$\frac{1}{16}$}
      \psframe[framearc=0,fillstyle=solid,fillcolor=red!20,linecolor=red!50](0,0)(3,1.47)
      \psframe[framearc=0,fillstyle=solid,fillcolor=red!20,linecolor=red!50](-1.5,0)(0,1.05)
      \psframe[framearc=0,fillstyle=solid,fillcolor=red!20,linecolor=red!50](-2.25,0)(-1.5,1.11)
      \psframe[framearc=0,fillstyle=solid,fillcolor=red!20,linecolor=red!50](-2.625,0)(-2.25,1.4)
      \psframe[framearc=0,fillstyle=solid,fillcolor=red!20,linecolor=red!50](-2.813,0)(-2.625,1.97)
      \psaxeslabels{->}(-3,0)(-3.5,-.5)(3.5,3){$x$}{$f_X(x)$}

      \psaxeslabels{->}(-3,-3)(-3.5,-3.5)(3.5,-0.5){$x$}{$g(x)$}
      \psTick{0}(-3,-1) \rput[th](-3,-1.1){\footnotesize$1$}
      \psline[style=Graph,linecolor=red](0,-3)(3,-1)\psdisk[fillcolor=red](3,-1){0.07}\pscircle[linecolor=red](0,-3){0.07}
      \psline[style=Graph,linecolor=red](-1.5,-3)(0,-1)\psdisk[fillcolor=red](0,-1){0.07}\pscircle[linecolor=red](-1.5,-3){0.07}
      \psline[style=Graph,linecolor=red](-2.25,-3)(-1.5,-1)\psdisk[fillcolor=red](-1.5,-1){0.07}\pscircle[linecolor=red](-2.25,-3){0.07}
      \psline[style=Graph,linecolor=red](-2.625,-3)(-2.25,-1)\psdisk[fillcolor=red](-2.25,-1){0.07}\pscircle[linecolor=red](-2.625,-3){0.07}
      \psline[style=Graph,linecolor=red](-2.813,-3)(-2.625,-1)\psdisk[fillcolor=red](-2.625,-1){0.07}\pscircle[linecolor=red](-2.813,-3){0.07}
    \end{pspicture}
  \caption{Piecewise bijective function and input density leading to infinite loss (cf.~Example~\ref{ex:infloss})}
  \label{fig:infLoss}
\end{figure}
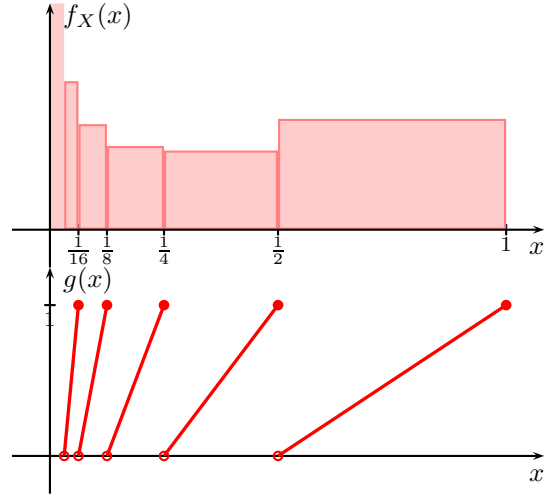

Assume that we consider the following scalar function $g{:}\ (0,1]\to(0,1]$, mapping every interval $(2^{-n},2^{-n+1}]$ onto the interval $(0,1]$:
\begin{equation}
 g(x) = 2^n (x-2^{-n}) \text{ if } x\in(2^{-n},2^{-n+1}],\ n\in\mathbb{N}
\end{equation}
Assume further that the PDF of the input $X$ is given as (see Fig.~\ref{fig:infLoss})
\begin{multline}
 f_X(x) = 2^n \left(\frac{1}{\log(n+1)}-\frac{1}{\log(n+2)}\right),\\\text{ if } x\in(2^{-n},2^{-n+1}],\ n\in\mathbb{N}.
\end{multline}
As an immediate consequence, the output RV $Y$ is uniformly distributed on $(0,1]$.

To apply Proposition~\ref{prop:partitionLoss}, we need
\begin{multline}
 \Prob{W=n|Y=y}= \Prob{W=n}\\=\frac{1}{\log(n+1)}-\frac{1}{\log(n+2)}.
\end{multline}
For this distribution, the entropy is known to be infinite~\cite{Baer_InfEntropy}, and thus
\begin{equation}
 \loss{X\to Y}=\ent{W|Y}=\ent{W}=\infty.
\end{equation}
\end{example}

\subsection{Upper Bounds on the Information Loss}\label{ssec:UBLoss}
The examples we examined so far were simple in the sense that the information loss could be computed in closed form. There are certainly cases where this is not possible, especially since the expressions involved in Proposition~\ref{prop:lossPBFdiffEnt} may involve a logarithm of a sum. It is therefore essential to accompany the exact expressions by bounds which are more simple to evaluate. In particular, we present a corollary to Proposition~\ref{prop:partitionLoss} which follows from the fact that conditioning reduces entropy:
\begin{cor}\label{cor:UBLossW}
 \begin{equation}
 \loss{X\to Y}\leq\ent{W}
\end{equation}
\end{cor}
Note that in both Example~\ref{ex:square} and~\ref{ex:infloss} this upper bound holds with equality.

The proof of Proposition~\ref{prop:partitionLoss} also allows us to derive the bounds presented in
\begin{prop}[Upper Bounds on Information Loss]\label{prop:UBLoss}
 The information loss induced by a PBF can be upper bounded by the following ordered set of inequalities:
\begin{IEEEeqnarray}{RCL}
 \loss{X\to Y} &\leq& \int_\dom{Y} f_Y(y) \log \card{\preim{y}} dy \label{eq:bound1}\\
 &\leq&  \log \left(\sum_i \int_{\dom{Y}_i} f_Y(y) dy \right)\label{eq:bound2} \\
&\leq& \mathrm{ess}\sup_{y\in\dom{Y}}\log\card{\preim{y}}\label{eq:bound2b} \\
&\leq& \log\card{\{\dom{X}_i\}}\label{eq:bound3}
\end{IEEEeqnarray}
where $\card{B}$ is the cardinality of the set $B$. Bound~\eqref{eq:bound1} holds with equality if and only if
\begin{equation}
 \sum_{x_k\in\preim{g(x)}} \frac{f_X(x_k)}{|\det\Jac{g}{x_k}|}\frac{|\det\Jac{g}{x}|}{f_X(x)}\stackrel{P_X\text{-a.s.}}{=}\card{\preim{g(x)}}. \label{eq:reqForBound}
\end{equation}
If and only if this expression is constant $P_X$-a.s., bounds~\eqref{eq:bound2} and~\eqref{eq:bound2b} are tight. Bound~\eqref{eq:bound3} holds with equality if and only if additionally $P_Y(\dom{Y}_i)=1$ for all $i$.
\end{prop}

\def \proofUBhere {0}
\begin{IEEEproof}
\ifthenelse{\proofUBhere=1}{The proof depends in parts on the proof of Proposition~\ref{prop:partitionLoss}, where we showed that
\begin{equation}
 \loss{X\to Y} = \int_\dom{Y} \ent{W|Y=y}f_Y(y) dy.
\end{equation}
The first inequality~\eqref{eq:bound1} is due to the maximum entropy property of the uniform distribution, i.e., $\ent{W|Y=y}\leq\log\card{\preim{y}}$ with equality if and only if $p(i|y)=1/\card{\preim{y}}$ for all $i$ for which $g_i^{-1}(y)\neq \emptyset$. But this translates to
\begin{equation}
 \card{\preim{y}}=\frac{|\det\Jac{g}{g_i^{-1}(y)}|f_Y(y)}{f_X(g_i^{-1}(y))}.
\end{equation}
Inserting the expression for $f_Y$ and substituting $x$ for $g_i^{-1}(y)$ (it is immaterial which $i$ is chosen, as long as the preimage of $y$ is not the empty set) we obtain
\begin{equation}
 \card{\preim{g(x)}}=\sum_{x_k\in\preim{g(x)}}\frac{f_X(x_k)}{|\det\Jac{g}{x_k}|}\frac{|\det\Jac{g}{x}|}{f_X(x)}.
\end{equation}

The second inequality~\eqref{eq:bound2} is due to Jensen~\cite[2.6.2]{Cover_Information2}, where we wrote
\begin{multline}
 \expec{\log\card{\preim{Y}}} \leq \log\expec{\card{\preim{Y}}}\\
=\log\int_\dom{Y} \card{\preim{y}} dP_Y(y)\\
=\log\int_\dom{Y} \sum_i \card{g_i^{-1}(y)} dP_Y(y)
=\log\sum_i\int_{\dom{Y}_i} dP_Y(y)
\end{multline}
since $\card{g_i^{-1}(y)}=1$ if $y\in\dom{Y}_i$ and zero otherwise. Equality is achieved if and only if $\card{\preim{y}}$ is constant $P_Y$-a.s. In this case also the third inequality~\eqref{eq:bound2b} is tight, which is obtained by replacing the expected value of the cardinality of the preimage by its essential supremum.

Finally, the cardinality of the preimage cannot be larger than the cardinality of the partition used in Definition~\ref{def:PBF}, which yields the last inequality~\eqref{eq:bound3}. For equality, consider that, assuming that all previous requirements for equality in the other bounds are fulfilled,~\eqref{eq:bound2} yields $\card{\{\dom{X}_i\}}$ if and only if $P_Y(\dom{Y}_i)=1$ for all $i$. This completes the proof.
\ifthenelse{\proofUBhere=1}{}{\endproof}
}{See Appendix~\ref{app:proofUB}.}
\end{IEEEproof}

Note that all bounds of Proposition~\ref{prop:UBLoss} hold with equality in Examples~\ref{ex:square} and~\ref{ex:infloss}. Clearly, examples where the PDF of $X$ and the absolute value of the Jacobian determinant are constant on $\dom{X}$ render the first bound~\eqref{eq:bound1} tight (cf.~\cite[conference version,~Sect.~VI]{Geiger_ILStatic_IZS}). Two other types of scenarios, where these bounds can hold with equality, are worth mentioning: First, for functions $g{:}\ \mathbb{R}\to\mathbb{R}$ equality holds if the function is related to the cumulative distribution function of the input RV such that, for all $x$, $|g'(x)|=f_X(x)$ (see extended version of~\cite{Geiger_ISIT2011arXiv}). The second case occurs when both function and PDF are ``repetitive'', in the sense that their behavior on $\dom{X}_1$ is copied to all other $\dom{X}_i$, and that, thus, $f_X(x_i)$ and $|\det\Jac{g}{x_i}|$ is the same for all elements of the preimage $\preim{y}$. Example~\ref{ex:square} represents such a case.

\subsection{Reconstruction and Reconstruction Error Probability}\label{ssec:reconLoss}
We now investigate connections between the information lost in a system and the probability for correctly reconstructing the system input. In particular, we present a series of Fano-type inequalities between the information loss and the reconstruction error probability. This connection is sensible, since the preimage of every output value is an at most countable set. 

Intuitively, one would expect that the fidelity of a reconstruction of a continuous input RV is best measured by some distance measure ``natural'' to the set $\dom{X}$, such as, e.g., the mean absolute distance or the mean squared-error (MSE), if $\dom{X}$ is a subset of the Euclidean space. However, as the following example shows, there is no connection between the information loss and such distance measures:

\begin{example}{Energy and Information behave differently}\label{ex:noMSELoss}
\begin{figure}[t]
 \centering
    \begin{pspicture}[showgrid=false](-4,-6)(4,2)
      \psaxeslabels{->}(0,0)(-4,-1.5)(4,2){$x$, \textcolor{red}{$\rec{y}$}}{$g_1(x)$, \textcolor{red}{$y$}}
      \psline[linecolor=red,linewidth=2.5pt](-0.75,-0.75)(0.75,0.75)  \psline(1.5,0.75)(0.75,0) \psline(-1.5,-0.75)(-0.75,0) \psline[linecolor=red,linewidth=2.5pt](-1.5,-0.75)(-2.25,-1.5) \psline[linecolor=red,linewidth=2.5pt](1.5,0.75)(2.25,1.5)
       \psline(-2.25,-0.75)(-3,-1.5) \psline(2.25,0.75)(3,1.5)
      \psTick[angle=90](0.75,0) \rput[tc](0.75,-0.2){\footnotesize$q_1$}
      \psTick[angle=90](3,0) \rput[tc](3,-0.2){\footnotesize$a$}\psTick[angle=90](-3,0) \rput[tc](-3,-0.2){\footnotesize$-a$}

      \psaxeslabels{->}(0,-4)(-4,-5.5)(4,-2){$x$, \textcolor{red}{$\rec{y}$}}{$g_2(x)$, \textcolor{red}{$y$}}
      \psline[linecolor=red,linewidth=2.5pt](-1.5,-5.5)(1.5,-2.5) \psline(-3,-5.5)(-1.5,-4) \psline(3,-2.5)(1.5,-4)
      \psTick[angle=90](1.5,-4) \rput[tc](1.5,-4.2){\footnotesize$q_2$}
      \psTick[angle=90](3,-4) \rput[tc](3,-4.2){\footnotesize$a$}\psTick[angle=90](-3,-4) \rput[tc](-3,-4.2){\footnotesize$-a$}
    \end{pspicture}
\caption{Two different functions $g_1$ and $g_2$ with the same information loss, but with a different mean-squared reconstruction error. The corresponding reconstructors are indicated by thick red lines. (It is immaterial whether the functions are left- or right-continuous.)}
\label{fig:noMSELoss}
\end{figure}
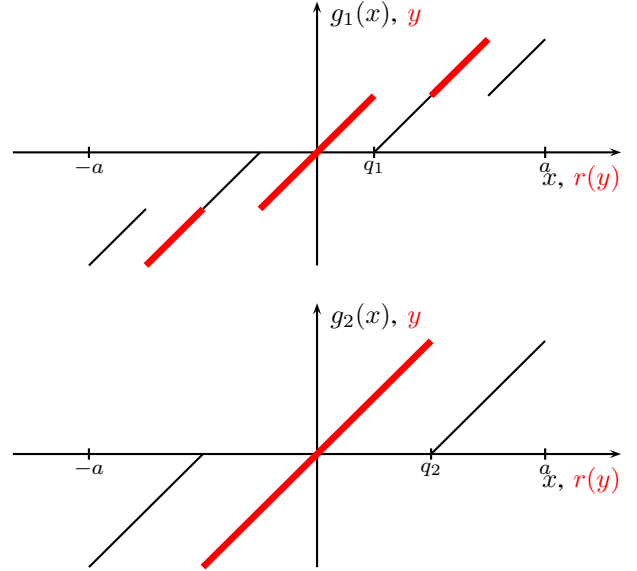
Consider the two functions $g_1$ and $g_2$ depicted in Fig.~\ref{fig:noMSELoss}, together with a possible reconstructor (see below). Assume further that the input RV is uniformly distributed on $[-a,a]$. It follows that
\begin{equation}
 \loss{X\to g_1(X)}=\loss{X\to g_2(X)} = 1
\end{equation}
The mean-squared reconstruction error $\expec{(X-\rec{Y})^2}$, however, differs for $Y=g_1(X)$ and $Y=g_2(X)$, since for $g_1$ the Euclidean distance between $x$ and $\rec{y}$ is generally smaller. In particular, decreasing the value of $q_1$ even further and extending the function accordingly would allow us to make the mean-squared reconstruction error arbitrarily small, while the information loss remains unchanged.
\end{example}

Clearly, a reconstructor trying to minimize the mean-squared reconstruction error will look totally different than a reconstructor trying to recover the input signal with high probability. Since for piecewise bijective functions a recovery of $X$ is possible (in contrast to noisy systems, where this is not the case), in our opinion a thorough analysis of such reconstructors is in order.

Aside from being of theoretical interest, there are practical reasons to justify the investigation: As already mentioned in Section~\ref{ssec:UBLoss}, the information loss is a quantity which is not always computable in closed form. If one can thus define a (sub-optimal) reconstruction of the input of the output for which the probability $\perr$ of error is easy to calculate, the Fano-type bounds would yield yet another set of upper bounds on the information loss. But also the reverse direction is of practical interest: Given the information loss $\loss{X\to Y}$ of a system, the presented inequalities allow one to bound the reconstruction error $\perr$. For example, one might want to obtain performance bounds of a non-coherent communications receiver (e.g., energy detector) in a semi-coherent broadcast scenario (e.g, combining pulse-position modulation and phase-shift keying, as in the IEEE 802.15.4a standard~\cite{4aStandard}).

We therefore present Fano-type inequalities to bound the reconstruction error probability via information loss. Due to the peculiarities of entropy pointed out in~\cite{Ho_Discont}, however, we restrict ourselves to finite partitions $\{\dom{X}_i\}$, guaranteeing a finite preimage for every output value. We note in passing that the results derived in this subsection not only apply to the mitigation of non-injective effects of deterministic systems, but to any reconstruction scenario where the cardinality of the input alphabet depends on the actual output value.

We start with introducing
\begin{definition}[Reconstructor \& Reconstruction Error]\label{def:reconstruction}
Let $r{:}\ \dom{Y}\to\dom{X}$ be a \emph{reconstructor}. Let $E$ denote the event of a \emph{reconstruction error}, i.e.,
\begin{equation}
 E=\begin{cases}
    1, & \text{ if } \rec{Y} \neq X\\ 0, & \text{ if } \rec{Y} = X
   \end{cases}.
\end{equation}
The probability of a reconstruction error is given by
\begin{equation}
 \perr = \Prob{E=1} = \int_{\dom{Y}} \perr(y) dP_Y(y)
\end{equation}
where $\perr(y)=\Prob{E=1|Y=y}$.
\end{definition}

In the following, we will investigate two different types of reconstructors: the maximum a-posteriori (MAP) reconstructor and a sub-optimal reconstructor. The MAP reconstructor chooses the reconstruction such that its conditional probability given the output is maximized, i.e., 
\begin{equation}
 \recmap{y} = \arg \max_{x_k\in\preim{y}} \Prob{X=x_k|Y=y}.
\end{equation}
In other words, with Definition~\ref{def:reconstruction} the MAP reconstructor minimizes $\perr(y)$. Interestingly, this reconstructor has a simple description for the problem at hand:

\begin{prop}[MAP Reconstructor]\label{prop:MAPRecon}
The MAP estimator for a PBF is
\begin{equation}
 \recmap{y} = g_k^{-1}(y)
\end{equation}
where
\begin{equation}
 k = \arg \max_{i:g_i^{-1}(y)\neq \emptyset} \left\{ \frac{f_X(g_i^{-1}(y))}{|\det\Jac{g}{g_i^{-1}(y)}|}\right\}.
\end{equation}
\end{prop}
\begin{IEEEproof}
 The proof follows from Corollary~\ref{cor:partitionLoss}, which states that, given $Y=y$ is known, reconstructing the input essentially amounts to reconstructing the partition from which it was chosen. The MAP reconstructor thus can be rewritten as
\begin{equation}
 \recmap{y} = \arg \max_{i} p(i|y).
\end{equation}
where we used the notation from the proof of Proposition~\ref{prop:partitionLoss}. From there,
\begin{equation}
 p(i|y)=\begin{cases}
     \frac{f_X(g_i^{-1}(y))}{|\det\Jac{g}{g_i^{-1}(y)}|f_Y(y)}, & \text{ if } g_i^{-1}(y) \neq \emptyset\\
0,&\text{ if } g_i^{-1}(y) = \emptyset
    \end{cases}
\end{equation}
This completes the proof.
\end{IEEEproof}

We derive Fano-type bounds for the MAP reconstructor, or any reconstructor for which $\rec{y}\in\preim{y}$. Under the assumption of a finite partition $\{\dom{X}_i\}$, note that Fano's inequality~\cite[pp.~39]{Cover_Information2}, where $\binent{p}=-p\log p-(1-p)\log(1-p)$,
\begin{equation}
 \loss{X\to Y} \leq \binent{\perr}+\perr\log\left(\card{\{\dom{X}_i\}}-1\right)\label{eq:FanoOriginal}
\end{equation}
trivially holds. We further note that in the equation above one can exchange $\card{\{\dom{X}_i\}}$ by $\mathrm{ess}\sup_{y\in\dom{Y}}\card{\preim{y}}$ to improve the bound. In what follows, we aim at further improvements.

\begin{definition}[Bijective Part]\label{def:biject}
 Let $\dom{X}_b$ be the maximal set such that $g$ restricted to this set is injective, and let $\dom{Y}_b$ be the image of this set. Thus, $g{:}\ \dom{X}_b\to\dom{Y}_b$ bijectively, where 
\begin{equation}
 \dom{X}_b=\{x\in\dom{X}{:}\ \card{\preim{g(x)}}=1\}.
\end{equation}
Then $P_b=P_X(\dom{X}_b)=P_Y(\dom{Y}_b)$ denotes the bijectively mapped probability mass.
\end{definition}

\begin{prop}[Fano-Type Bound]\label{prop:fanoLBLoss}
For the MAP reconstructor -- or any reconstructor for which $\rec{y}\in\preim{y}$ -- the information loss $\loss{X\to Y}$ is upper bounded by
\begin{multline}
 \loss{X\to Y} \leq\min\{1-P_b, \binent{\perr}\}-\perr\log\perr\\+\perr\log\left(\expec{\card{\preim{Y}}-1}\right).\label{eq:fano3}
\end{multline}
\end{prop}

\def \proofMAPFanohere {0}
\begin{IEEEproof}
\ifthenelse{\proofMAPFanohere=1}{\input{pbfs_MAPFano.tex}}{See Appendix~\ref{app:proofMAPFano}.}
\end{IEEEproof}

If we compare this result with Fano's original bound~\eqref{eq:FanoOriginal}, we see that the cardinality of the partition is replaced by the expected cardinality of the preimage. Due to the additional term $\perr\log\perr$ this improvement is only potential, since there exist cases where Fano's original bound is better. An example is the square-law device of Example~\ref{ex:square}, for which Fano's inequality is tight, but for which Proposition~\ref{prop:fanoLBLoss} would yield $\loss{X\to Y}\leq 2$.

For completeness, we want to mention that for the MAP reconstructor also a lower bound on the information loss can be given. We restate
\begin{prop}[Feder \& Merhav,~\cite{Feder_EntropyError}]\label{prop:federUBLoss}
 The information loss $\loss{X\to Y}$ is lower bounded by the error probability $\perr$ of a MAP reconstructor by
\begin{equation}
 \phi(\perr) \leq \loss{X\to Y}
\end{equation}
where $\phi(x)$ is a piecewise linear function defined as
\begin{equation}
 \phi(x) = \left(x-\frac{i-1}{i}\right)(i+1)i\log \left(1+\frac{1}{i}\right)+\log i\label{eq:piecewisebound}
\end{equation}
for $\frac{i-1}{i} \leq x \leq \frac{i}{i+1}$.
\end{prop}
At the time of submission we were not able to improve this bound for the present context since the cardinality of the preimage has no influence on $\phi$.

We illustrate the utility of these bounds, together with those presented in Proposition~\ref{prop:UBLoss}, in

\begin{example}{Third-order Polynomial}\label{ex:3rd}
 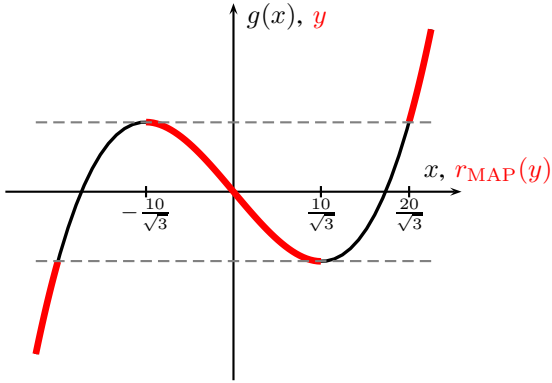
\begin{figure}[t]
 \centering
	\begin{pspicture}[showgrid=false](-4,-2.5)(4,3)
		\psaxeslabels[xlpos=t]{->}(0,0)(-3,-2.5)(3,2.5){}{$g(x)$, \textcolor{red}{$y$}}
		\put(2.5,0.2){$x$, \textcolor{red}{$\recmap{y}$}}
		\psplot[style=Graph]{-2.6}{2.6}{x x mul x mul x -4 mul add 0.3 mul}
		\psplot[style=Graph,linecolor=red,linewidth=2.5pt]{-2.6}{-2.31}{x x mul x mul x -4 mul add 0.3 mul}
		\psplot[style=Graph,linecolor=red,linewidth=2.5pt]{2.31}{2.6}{x x mul x mul x -4 mul add 0.3 mul}
		\psTick{90}(-1.15,0) \rput[th](-1.15,-0.1){\footnotesize$-\frac{10}{\sqrt{3}}$}
		\psTick{90}(1.15,0) \rput[th](1.15,-0.1){\footnotesize$\frac{10}{\sqrt{3}}$}
		\psTick{90}(2.31,0) \rput[th](2.31,-0.1){\footnotesize$\frac{20}{\sqrt{3}}$}
 		\psplot[style=Graph,linecolor=red,linewidth=2.5pt]{-1.15}{1.15}{x x mul x mul x -4 mul add 0.3 mul}
		\psline[style=Dash, linecolor=gray](-2.6,0.92)(2.6,0.92) \psline[style=Dash, linecolor=gray](-2.6,-0.92)(2.6,-0.92)
	\end{pspicture}
\caption{Third-order polynomial of Example~\ref{ex:3rd} and its MAP reconstructor indicated with a thick red line.}
\label{fig:3rd}
\end{figure}
Consider the function depicted in Fig.~\ref{fig:3rd}, which is defined as
\begin{equation}
 g(x) = x^3-100x.
\end{equation}
The input to this function is a zero-mean Gaussian RV $X$ with variance $\sigma^2$. A closed-form evaluation of the information loss is not possible, since the integral involves the logarithm of a sum. However, we note that
\begin{equation}
 \dom{X}_b = \left(-\infty,-\frac{20}{\sqrt{3}}\right] \cup \left[\frac{20}{\sqrt{3}},\infty\right)
\end{equation}
and thus $P_b=2Q\left(\frac{20}{\sqrt{3}\sigma}\right)$, where $Q$ denotes the $Q$-function~\cite[26.2.3]{Abramowitz_Handbook}. With a little algebra we thus obtain the bounds from Proposition~\ref{prop:UBLoss} as
\begin{equation}
 \loss{X\to Y} \leq (1-P_b)\log 3\leq \log\left(3-2P_b\right)\leq \log 3
\end{equation}
where $\mathrm{ess}\sup_{y\in\dom{Y}}\card{\preim{y}}=\card{\{\dom{X}_i\}}=3$.

As it can be shown\footnote{The authors thank Stefan Wakolbinger for pointing us to this fact.}, the MAP reconstructor assumes the properties depicted in Fig.~\ref{fig:3rd}, from which an error probability of
\begin{equation}
 \perr = 2Q\left(\frac{10}{\sqrt{3}\sigma}\right)-2Q\left(\frac{20}{\sqrt{3}\sigma}\right)
\end{equation}
can be computed. We display Fano's bound together with the bounds from Propositions~\ref{prop:UBLoss},~\ref{prop:fanoLBLoss}, and~\ref{prop:federUBLoss} in Fig.~\ref{fig:3rdLoss}
\begin{figure}[t]
 \centering
\begin{pspicture}[showgrid=false](-0.5,-0.5)(8,6)
	\footnotesize
	\psaxes[Dx=10,dx=1.5,Dy=0.5,dy=1.25]{->}(0,0)(7.5,5.5)[$\sigma$,-90][$\loss{X\to Y}$,0]
	\rput[rt](7.5,5.5){\psframebox%
	{\begin{tabular}{ll}
		\psline[linewidth=1pt,linecolor=black](0.1,0.1)(0.7,0.1)&\hspace*{0.5cm}  Numeric Result\\
	  	\psline[linewidth=1pt,linecolor=red](0.1,0.1)(0.7,0.1) &\hspace*{0.5cm}  Upper Bound (Prop.~\ref{prop:UBLoss})\\%
		\psline[linewidth=1pt,linecolor=blue,style=Dash](0.1,0.1)(0.7,0.1) &\hspace*{0.5cm}  Fano's Bound\\%
		\psline[linewidth=1pt,linecolor=blue](0.1,0.1)(0.7,0.1) &\hspace*{0.5cm}  Fano-type Bound (Prop.~\ref{prop:fanoLBLoss})\\
		\psline[linewidth=1pt,linecolor=magenta,style=Dash](0.1,0.1)(0.7,0.1) &\hspace*{0.5cm}  Lower Bound (Prop.~\ref{prop:federUBLoss})
	 \end{tabular}}}
	\readdata{\Num}{matlab/3rdNumeric.dat}
	\readdata{\Fano}{matlab/3rdFano.dat}
	\readdata{\FanoType}{matlab/3rdFanoType.dat}
	\readdata{\UB}{matlab/3rdUB.dat}
	\readdata{\LB}{matlab/3rdLB.dat}
	\psset{xunit=0.15cm,yunit=2.5cm}
	 	\dataplot[plotstyle=curve,linecolor=black,linewidth=1pt]{\Num}
		\dataplot[plotstyle=curve,linecolor=blue,style=Dash,linewidth=0.5pt]{\Fano}
		\dataplot[plotstyle=curve,linecolor=blue]{\FanoType}
		\dataplot[plotstyle=curve,linecolor=magenta,style=Dash]{\LB}
		\dataplot[plotstyle=curve,linecolor=red]{\UB}
\end{pspicture}
\caption{Information loss for Example~\ref{ex:3rd} as a function of input variance $\sigma^2$.}
\label{fig:3rdLoss}
\end{figure}
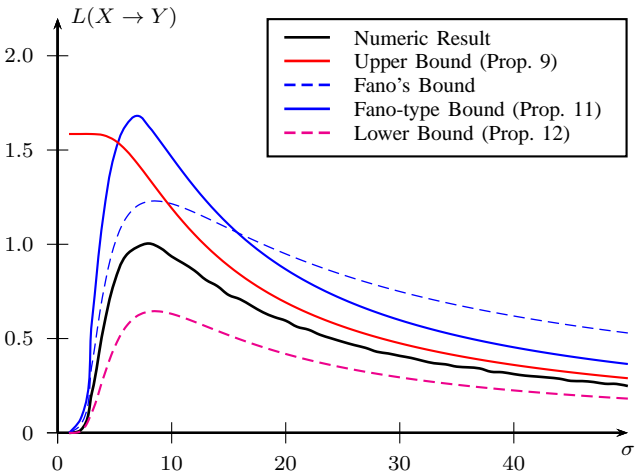
\end{example}

What becomes apparent from this example is that the bounds from Propositions~\ref{prop:UBLoss} and~\ref{prop:fanoLBLoss} cannot form an ordered set; the same holds for Fano's inequality, which can be better or worse than our Fano-type bound, depending on the scenario.

While in this example the MAP reconstructor was relatively simple to find, this might not always be the case. For bounding the information loss of a system (rather than reconstructing the input), it is therefore desirable to introduce a simpler, sub-optimal reconstructor:
\newcommand{\recsub}[1]{r_{\mathrm{sub}}(#1)}
\begin{prop}[Suboptimal Reconstruction]\label{prop:Suboptimal}
 Consider the following sub-optimal reconstructor
\begin{equation}
 \recsub{y} = \begin{cases}
   g^{-1}(y), & \text{if }y\in\dom{Y}_b\\
 g_k^{-1}(y), & \text{if }y\in\dom{Y}_k\setminus\dom{Y}_b\\
 x{:}\ x\in\dom{X}_k, & \text{else }
             \end{cases}
\end{equation}
where
\begin{equation}
 k=\arg \max_i P_X(\dom{X}_i\cup\dom{X}_b)
\end{equation}
and where $\dom{Y}_k=g(\dom{X}_k)$. 

Letting $\overline{K}=\mathrm{ess}\sup_{y\in\dom{Y}}\card{\preim{y}}$ and with the error probability 
\begin{eqnarray}
 \perh=1-P_X(\dom{X}_k\cup\dom{X}_b)
\end{eqnarray}
of this reconstructor, the information loss is upper bounded by the following, Fano-type inequality:
\begin{equation}
 \loss{X\to Y}\leq 1-P_b+\perh\log\left( \overline{K}-1\right)
\end{equation}
\end{prop}

\begin{IEEEproof}
 See~Appendix~\ref{app:proofFano}
\end{IEEEproof}

This reconstructor is \emph{simple} in the sense that the reconstruction is always chosen from the element $\dom{X}_k$ containing most of the probability mass, after considering the set on which the function is bijective. This allows for a simple evaluation of the reconstruction error probability $\perh$, which is independent of the Jacobian determinant of $g$.

It is interesting to see that the Fano-type bound derived here permits a similar expression as derived in Proposition~\ref{prop:fanoLBLoss}, despite the fact that the sub-optimal reconstructor not necessarily satisfies $\recsub{y}\in\preim{y}$. For this type of reconstructors, $(\card{\cdot}-1)$ typically has to be replaced by $\card{\cdot}$. We thus note that also the following bounds hold:
\begin{IEEEeqnarray}{RCL}
 \loss{X\to Y} &\leq& \binent{\perh}+\perh\log\left(\mathrm{ess}\sup_{y\in\dom{Y}}\card{\preim{y}}\right)\\
&\leq& \binent{\perh}+\perh\log\left(\card{\{\dom{X}_i\}}\right)\\
 \loss{X\to Y} &\leq&\min\{1-P_b, \binent{\perr}\}-\perr\log\perr\notag \\&&{}+\perr\log\left(\expec{ \card{\preim{Y}}}\right)
\end{IEEEeqnarray}

Before proceeding, we want to briefly reconsider

\begin{example}[ex:infloss]{Infinite Loss (revisited)}
 For the PDF and the function depicted in Fig.~\ref{fig:infLoss} it was shown that the information loss was infinite. By recognizing that the probability mass contained in $\dom{X}_1=(\frac{1}{2},1]$ exceeds the mass contained in all other subsets, we obtain an error probability for reconstruction equal to
\begin{equation}
 \perh = \frac{1}{\log 3} \approx 0.63.
\end{equation}
In this particular case we even have $\perr=\perh$, since the MAP reconstructor coincides with the suboptimal reconstructor. Since in this case $\card{\preim{y}}=\infty$ for all $y\in\dom{Y}$, all upper bounds derived from Fano-type bounds evaluate to infinity.
\end{example}

\section{Information Loss for Functions which Reduce Dimensionality}\label{sec:uncoutPreim}
We now analyze systems for which the absolute information loss $\loss{X\to Y}$ is infinite. Aside from practically irrelevant cases as in Example~\ref{ex:infloss}, this subsumes cases where the dimensionality of the input signal is reduced, e.g., by dropping coordinates or by keeping the function constant on a subset of its domain.

Throughout this section we assume that the input RV has positive information dimension, i.e., $\infodim{X}>0$ and infinite entropy $\ent{X}=\infty$. We further assume that the function $g$ describing the system is such that the \emph{relative} information loss $\relLoss{X\to Y}$ is positive (from which $\loss{X\to Y}=\infty$ follows; cf.~Proposition~\ref{prop:RelInfLoss}).

According to Proposition~\ref{prop:dimTrans} the relative information loss in a Lipschitz function is positive whenever the \emph{information dimension} of the input RV $X$ is reduced. Interestingly, a reduction of the dimension of the support $\dom{X}$ does not necessarily lead to a positive relative information loss, nor does its preservation guarantee vanishing relative information loss. 

\subsection{Relative Information Loss for Continuous Input RVs}\label{ssec:RILdimRed}
We again assume that $\dom{X}\subseteq\mathbb{R}^N$ and $P_X\ll\mu^N$, thus $\infodim{X}=N$. We already found in Proposition~\ref{prop:dimTrans} that for Lipschitz functions the relative information loss is given as
\begin{equation}
 \relLoss{X\to Y}=1-\frac{\infodim{Y}}{N}
\end{equation}
where $Y$ may be a mixture of RVs with different information dimensions (for which $\infodim{Y}$ can be computed; cf.~\cite{Wu_Renyi,Smieja_EntropyMixture}). Such a mixture may result, e.g., from a function $g$ mapping different subsets of $\dom{X}$ to sets of different covering dimension. We intend to make this statement precise in what follows.

First, let us drop the requirement of Lipschitz continuity; generally, we now cannot expect Proposition~\ref{prop:dimTrans} to hold. We assume that $g$ is piecewise defined, as in Definition~\ref{def:PBF}. Here, however, we do not require $g_i{:}\ \dom{X}_i\to\dom{Y}_i$ to be bijective, but to be a \emph{submersion}, i.e., a smooth function between smooth manifolds whose pushforward is surjective everywhere (see, e.g.,~\cite{Lee_SmoothManifolds}). A projection onto any $M<N$ coordinates of $X$, for example, is a submersion. With these things in mind, we present

\begin{prop}[Relative Information Loss in Dimensionality Reduction]\label{prop:RILdimRed}
  Let $\{\dom{X}_i\}$ be a partition of $\dom{X}$ such that each of its $K$ elements is a smooth $N$-dimensional manifold. Let $g$ be such that $g_i=g|_{\dom{X}_i}$ are submersions to smooth $M_i$-dimensional manifolds $\dom{Y}_i$ ($M_i\leq N$). Then, the relative information loss is
\begin{equation}
 \relLoss{X\to Y} = \sum_{i=1}^K P_X(\dom{X}_i)\frac{N-M_i}{N}.
\end{equation}
\end{prop}

\def \proofRILdimRed {0}
\begin{IEEEproof}
\ifthenelse{\proofRILdimRed=1}{ We start by noting that by the submersion theorem~\cite[Cor.~5.25]{Lee_SmoothManifolds} for any point $y\in\dom{Y}=\bigcup_{i=1}^K\dom{Y}_i$ the preimage under $g|_{\dom{X}_i}$ is either the empty set (if $y\notin\dom{Y}_i$) or an $(N-M_i)$-dimensional embedded submanifold of $\dom{X}_i$. We now write with~\cite{Wu_Renyi,Smieja_EntropyMixture}
\begin{multline}
 \infodim{X|Y=y}=\sum_{i=1}^K \infodim{X|Y=y,X\in\dom{X}_i}P_{X|Y=y}(\dom{X}_i)\\
 \leq\sum_{i=1}^K(N-M_i) P_{X|Y=y}(\dom{X}_i)
\end{multline}
since the information dimension of an RV cannot exceed the covering dimension of its support. Taking the expectation w.r.t. $Y$ yields
\begin{multline}
 \infodim{X|Y}\leq\sum_{i=1}^K(N-M_i) \int_\dom{Y} P_{X|Y=y}(\dom{X}_i)dP_Y(y)\\
=\sum_{i=1}^K(N-M_i)P_{X}(\dom{X}_i)
\end{multline}
and thus
\begin{equation}
 \relLoss{X\to Y} \leq \sum_{i=1}^K\frac{N-M_i}{N}P_{X}(\dom{X}_i)
\end{equation}
by Proposition~\ref{prop:RILDim}. It remains to show the reverse inequality.

To this end, note that without the Lipschitz condition Proposition~\ref{prop:dimTrans} would read
\begin{equation}
 \relTrans{X\to Y}\leq \frac{\infodim{Y}}{\infodim{X}}.
\end{equation}
But with~\cite{Wu_Renyi,Smieja_EntropyMixture} we can write for the information dimension of $Y$
\begin{equation}
 \infodim{Y} = \sum_{i=1}^K \infodim{Y|X\in\dom{X}_i}P_{X}(\dom{X}_i).
\end{equation}
By the fact that $g_i$ are submersions, preimages of $\mu^{M_i}$-null sets are $\mu^N$-null sets~\cite{Ponomarev_Preimage} -- were there a $\mu^{M_i}$-null set $B\subseteq\dom{Y}_i$ such that the conditional probability measure $P_{Y|X\in\dom{X}_i}(B)>0$, there would be some  $\mu^N$-null set $g_i^{-1}(B)\subseteq\dom{X}_i$ such that $P_{X|X\in\dom{X}_i}(g_i^{-1}(B))>0$, contradicting $P_X\ll\mu^N$. Thus, $P_{Y|X\in\dom{X}_i}\ll\mu^{M_i}$ and we get
\begin{equation}
 \relLoss{X\to Y} \geq 1-\sum_{i=1}^K \frac{M_i}{N}P_{X}(\dom{X}_i)=\sum_{i=1}^K\frac{N-M_i}{N}P_{X}(\dom{X}_i).
\end{equation}
This proves the reverse inequality and completes the proof.
\ifthenelse{\proofRILdimRed=1}{}{\endproof}}{See Appendix~\ref{app:proofRILdimRed}.}
\end{IEEEproof}

This result shows that the statement of Proposition~\ref{prop:dimTrans} not only holds for Lipschitz functions $g$, but for a larger class of systems yet to be identified.

We present two Corollaries to Proposition~\ref{prop:RILdimRed} concerning projections onto a subset of coordinates and functions which are constant on some subset with positive $P_X$-measure.

\begin{cor}\label{cor:RILProj}
Let $g$ be any projection of $X$ onto $M$ of its coordinates. Then, the relative information loss is
\begin{equation}
 \relLoss{X\to Y} = \frac{N-M}{N}.
\end{equation}
\end{cor}

\begin{cor}\label{cor:RILConst}
Let $g$ be constant on a set $A\subseteq\dom{X}$ with positive $P_X$-measure. Let furthermore $g$ be such that $\card{\preim{y}}<\infty$ for all $y\notin g(A)$. Then, the relative information loss is
\begin{equation}
 \relLoss{X\to Y} = P_X(A).
\end{equation}
\end{cor}

The first of these two corollaries has been applied to principle component analysis in~\cite{Geiger_RILPCA_arXiv}, while the second allows us to take up Example~\ref{ex:cclipper} (center clipper) again: There, we showed that both the information loss and the information transfer are infinite. For the relative information loss we can now show that it corresponds to the probability mass contained in the clipping region, i.e., $\relLoss{X\to Y}=P_X([-c,c])$.

The somewhat surprising consequence of these results is that the shape of the PDF has no influence on the relative information loss;  whether the PDF is peaky in the clipping region or flat, or whether the omitted coordinates are highly correlated to the preserved ones does neither increase nor decrease the relative information loss.

In particular, in~\cite{Geiger_RILPCA_arXiv} we showed that dimensionality reduction after performing a principle component analysis  leads to the same relative information loss as directly dropping $N-M$ coordinates of the input vector $X$.

\subsection{Bounds on the Relative Information Loss}\label{ssec:RILBounds}
Complementing the results from Section~\ref{ssec:UBLoss} we now present bounds on the relative information loss for some particular cases. We note in passing that from the trivial bounds on the information dimension ($\infodim{X}\in[0,N]$ if $\dom{X}$ is a subset of the $N$-dimensional Euclidean space or a sufficiently smooth $N$-dimensional manifold) simple bounds on the relative information loss can be computed. 

Here we present bounds on the relative information transfer and the relative information loss for an $N$-dimensional input RV by the corresponding coordinate-wise quantities.

\newcommand{\Y}[1]{Y^{(#1)}}
\newcommand{\Yh}[1]{\hat{Y}^{(#1)}}
\newcommand{\X}[1]{X^{(#1)}}
\newcommand{\Xh}[1]{\hat{X}^{(#1)}}
\begin{prop}[Upper Bound on the Relative Information Transfer]\label{prop:UBRelTrans}
 Let $g$ be a Lipschitz function with $N$-dimensional input $X$ and $K$-dimensional output $Y$. The relative information transfer is bounded by
\begin{equation}
\relTrans{X\to Y} \leq \sum_{i=1}^K\relTrans{X\to \Y{i}}
\end{equation}
where $\Y{i}$ is the $i$-th coordinate of $Y$.
\end{prop}

\def \proofUBRelTrans {0}
\begin{IEEEproof}
\ifthenelse{\proofUBRelTrans=1}{ The proof follows from Proposition~\ref{prop:dimTrans}, Definition~\ref{def:infodim}, and the fact that conditioning reduces entropy. We write
\begin{align}
 \relTrans{X\to Y} &= \relTrans{X\to \Y{1}, \Y{2},\dots,\Y{K}}   \\&= \frac{\infodim{\Y{1}, \Y{2},\dots,\Y{K}}}{\infodim{X}}\\
&=\limn\frac{\ent{\Yh{1}_n,\dots,\Yh{K}_n}}{\ent{\hat{X}_n}}\\&=\limn\frac{\sum_{i=1}^K \ent{\Yh{i}_n|\Yh{1}_n,\dots,\Yh{i-1}_n}}{\ent{\hat{X}_n}}\\
&\leq \limn\frac{\sum_{i=1}^K \ent{\Yh{i}_n}}{\ent{\hat{X}_n}} = \frac{\sum_{i=1}^K \infodim{\Yh{i}}}{\infodim{X}}\\&=\sum_{i=1}^K\relTrans{X\to \Y{i}}
\end{align}
where we exchanged limit and summation by the same reason as in the proof of Proposition~\ref{prop:RILDim}.
\ifthenelse{\proofUBRelTrans=1}{}{\endproof}}{See Appendix~\ref{app:proofUBRelTrans}.}
\end{IEEEproof}

This upper bound on the relative information transfer (which leads to a lower bound on the relative information loss) can also be applied if the system has $K$ one-dimensional output RVs, in which case $Y$ denotes denotes their collection.

\begin{prop}[Upper Bound on the Relative Information Loss]\label{prop:UBRelLoss}
 Let $X$ be an $N$-dimensional RV with a probability measure $P_X\ll\mu^N$ and let $Y$ be $N$-dimensional. Then,
\begin{equation}
 \relLoss{X\to Y} \leq \frac{1}{N}\sum_{i=1}^N\relLoss{\X{i}\to Y}\leq \frac{1}{N}\sum_{i=1}^N\relLoss{\X{i}\to \Y{i}}
\end{equation}
where $\X{i}$ and $\Y{i}$ are the $i$-th coordinates of $X$ and $Y$, respectively.
\end{prop}

\def \proofUBRelLoss {0}
\begin{IEEEproof}
\ifthenelse{\proofUBRelLoss=1}{ The proof follows from the fact that $\infodim{X}=N$ and, for all $i$, $\infodim{\X{i}}=1$. We obtain from the definition of relative information loss
\begin{multline}
 \relLoss{X\to Y} = \limn \frac{\ent{\Xh{1}_n,\dots,\Xh{N}_n|Y}}{\ent{\Xh{1}_n,\dots,\Xh{N}_n}}\\
= \limn \frac{\sum_{i=1}^N\ent{\Xh{i}_n|\Xh{1}_n,\dots,\Xh{i-1}_n,Y}}{\ent{\Xh{1}_n,\dots,\Xh{N}_n}}\\
\leq \limn\frac{\sum_{i=1}^N\ent{\Xh{i}_n|Y}}{\ent{\Xh{1}_n,\dots,\Xh{N}_n}}\label{eq:proofUBRelLoss}
\end{multline}
Exchanging again limit and summation we get
\begin{equation}
 \relLoss{X\to Y} \leq \frac{\sum_{i=1}^N \infodim{\X{i}|Y}}{\infodim{X}}=\frac{\sum_{i=1}^N \infodim{\X{i}|Y}}{N}.
\end{equation}
But since $N=N\infodim{\X{i}}$ for all $i$, we can write
\begin{equation}
 \relLoss{X\to Y} \leq \frac{1}{N} \sum_{i=1}^N \frac{\infodim{\X{i}|Y}}{\infodim{\X{i}}} = \frac{1}{N} \sum_{i=1}^N\relLoss{\X{i}\to Y}.
\end{equation}
This proves the first inequality. The second is obtained by removing conditioning again in~\eqref{eq:proofUBRelLoss}, since $Y=\{\Y{1},\dots,\Y{N}\}$.
\ifthenelse{\proofUBRelLoss=1}{}{\endproof}}{See Appendix~\ref{app:proofUBRelLoss}.}
\end{IEEEproof}


\begin{example}{Projection}\label{ex:projection}
 Let $X$ be an $N$-dimensional RV with probability measure $P_X\ll\mu^N$ and let $\X{i}$ denote the $i$-th coordinate of $X$. Let $g$ be a projection onto the first $M<N$ coordinates. The information loss is given as
\begin{equation}
 \relLoss{X\to Y}=\frac{N-M}{N}
\end{equation}
by Corollary~\ref{cor:RILProj}. Note further that $\relTrans{X\to\Y{i}} = \frac{1}{N}$ for all $i\in\{1,\dots,M\}$, which renders the bound of Proposition~\ref{prop:UBRelTrans} tight. Furthermore, $\relLoss{\X{i}\to Y}=0$ for $i\in\{1,\dots,M\}$, while $\relLoss{\X{i}\to Y}=1$ for $i\in\{M+1,\dots,N\}$ which shows tightness of Proposition~\ref{prop:UBRelLoss} as well.
\end{example}

\subsection{Reconstruction and Reconstruction Error Probability}\label{ssec:reconRIL}
We next take up the approach of Section~\ref{ssec:reconLoss} and present Fano-type relations between the relative information loss and the probability of a reconstruction error. While for piecewise bijective functions this relation was justified by the fact that for every output value the preimage under the system function is a countable set, the case is completely different here: Quantizers, for example, characterized with relative information loss in our framework, are typically evaluated based on some energetic measures (e.g., the mean-squared reconstruction error). As the following example shows, the relative information loss does not permit a meaningful interpretation in energetic terms, again underlining the intrinsically different behavior of information and energy measures.

\begin{example}[ex:quantizer]{Quantizer (revisited)}
 We now consider a continuous one-dimensional RV $X$ ($P_X\ll\mu$) and the quantizer introduced in Section~\ref{ssec:notation}. Since the quantizer is constant $P_X$-a.s., we obtain with Corollary~\ref{cor:RILConst}
\begin{equation}
 \relLoss{X\to \hat{X}_n} = 1.
\end{equation}
In other words, the quantizer destroys 100\% of the information available at its input. This naturally holds for all $n$, so a finer partition $\partit{n}$ cannot decrease the relative information loss. Conversely, the mean-squared reconstruction error decreases with increasing $n$.
\end{example}

We therefore turn to find connections between relative information loss and the reconstruction error probability after introducing
\newcommand{\boxdim}[1]{d_B(#1)}
\begin{definition}[Minkowski Dimension]\label{def:minkowski}
 The Minkowski- or box-counting dimension of a compact set $\dom{X}\subset\mathbb{R}^N$ is
\begin{equation}
 \boxdim{\dom{X}} = \limn \frac{\log\card{\partit{n}}}{n}
\end{equation}
where the partition $\partit{n}$ is induced by a uniform vector quantizer with quantization interval $\frac{1}{2^n}$.
\end{definition}

The Minkowski dimension of a set equals the information dimension of a uniform distribution on that set (e.g.,~\cite{Farmer_Dim}), and is a special case of R\'{e}nyi information dimension where the entropy is replaced with the R\'{e}nyi entropy of zeroth order~\cite{Grassberger_Dim}. We are now ready to state

\begin{prop}\label{prop:fanoUBRIL}
 Let $X$ be an RV with a probability measure $P_X$ with positive information dimension $\infodim{X}$ supported on a compact set $\dom{X}\subset\mathbb{R}^N$ with positive Minkowski dimension $\boxdim{\dom{X}}$. Then, the error probability bounds the relative information loss from above, i.e.,
\begin{equation}
 \relLoss{X\to Y}\leq\perr\frac{\boxdim{\dom{X}}}{\infodim{X}}.
\end{equation}
\end{prop}

\def \prooffanoUBRIL {0}
\begin{IEEEproof}
\ifthenelse{\prooffanoUBRIL=1}{Note that by the compactness of $\dom{X}$ the quantized input $\hat{X}_n$ has a finite alphabet, which allows us to employ Fano's inequality
\begin{equation}
 \ent{\hat{X}_n|Y} \leq \binent{P_{e,n}} + P_{e,n}\log\card{\partit{n}}
\end{equation}
where
\begin{equation}
 P_{e,n} = \Prob{\rec{Y}\neq\hat{X}_n}.
\end{equation}
Since Fano's inequality holds for arbitrary reconstructors, we let $\rec{\cdot}$ be the composition of the MAP reconstructor $\recmap{\cdot}$ and the quantizer introduced in Section~\ref{ssec:notation}. Consequently, $P_{e,n}$ is the probability that $\recmap{Y}$ and $X$ do not lie in the same quantization bin. Since the bin volumes reduce with increasing $n$, $P_{e,n}$ increases monotonically to $\perr$. We thus obtain with $\binent{p}\leq 1$ for $0\leq p \leq 1$
\begin{equation}
 \ent{\hat{X}_n|Y} \leq 1+ \perr\log\card{\partit{n}}.
\end{equation}
With the introduced definitions,
\begin{IEEEeqnarray}{RCL}
 \relLoss{X\to Y} &=& \limn \frac{\ent{\hat{X}_n|Y}}{\ent{\hat{X}_n}}\\
&\leq& \limn \frac{1+ \perr\log\card{\partit{n}}}{\ent{\hat{X}_n}}\\
&\stackrel{(a)}{=}&\perr\frac{\boxdim{\dom{X}}}{\infodim{X}}
\end{IEEEeqnarray}
where $(a)$ is obtained by dividing both numerator and denominator by $n$ and evaluating the limit. This completes the proof.
\ifthenelse{\prooffanoUBRIL=1}{}{\endproof}}{See Appendix~\ref{app:prooffanoUBRIL}.}
\end{IEEEproof}

In the case where $\dom{X}\subset\mathbb{R}^N$ and $P_X\ll\mu^N$ it can be shown\footnote{Always, $\infodim{X}\leq\boxdim{\dom{X}}\leq N$ if $\dom{X}\subset\mathbb{R}^N$, e.g., by~\cite[Thm.~1 and Lem.~4]{Wu_PhD}. $\infodim{X}=N$ leads to the desired result.} that this result simplifies to $\relLoss{X\to Y}\leq\perr$. Comparing this to the results of Example~\ref{ex:quantizer}, we can see that for a quantizer the reconstruction error probability is always $\perr=1$.

\begin{figure}[t]
  \centering
\begin{pspicture}[showgrid=false](-0.5,-0.5)(6,6)
  \psaxeslabels{->}(0,0)(-0.5,-0.5)(6,6){$\perr$}{$\relLoss{X\to Y}$}
  \psline[style=Graph,linecolor=red](0,0)(5,0)
  \psline[style=Graph,linecolor=red](5,0)(5,5)
  \psline[style=Graph,linecolor=red](3,5)(5,5)
  \psline[style=Graph,linecolor=red](3,5)(0,0)
  \psline[style=Dash,linecolor=gray](3,-0.2)(3,5)
  \psTick{90}(5,0) \rput[th](5,-0.1){\footnotesize$1$} \psTick{0}(0,5) \rput[th](-0.15,5){\footnotesize$1$}
  \psTick{90}(3,0) \rput[th](3,-0.3){\footnotesize$\infodim{X}/\boxdim{\dom{X}}$}

  \dotnode(3,5){n1}\uput[45](n1){\ref{ex:cclipper}}
  \dotnode(1.5,2.5){n3}\uput[45](n3){\ref{ex:cclipper}}
  \dotnode(2,2.5){n4}\uput[45](n4){\ref{ex:cclipper}}
  \dotnode(5,3){np}\uput[45](np){\ref{ex:projection}}
  \dotnode(5,5){n2}\uput[45](n2){\ref{ex:quantizer}}
  \dotnode(3.15,0){n5}\uput[45](n5){\ref{ex:infloss}}
  \dotnode(2.5,0){n8}\uput[45](n8){\ref{ex:square}}
\end{pspicture}
\caption{The accessible region for a $(\perr,\relLoss{X\to Y})$-pair for $\infodim{X}=0.6\boxdim{\dom{X}}$. Points with numbers indicate references to the numbered examples in this work. Note that the center clipper (example~\ref{ex:cclipper}) occurs three times in the plot, representing each instance in the text.}
\label{fig:pairregion}
\end{figure}
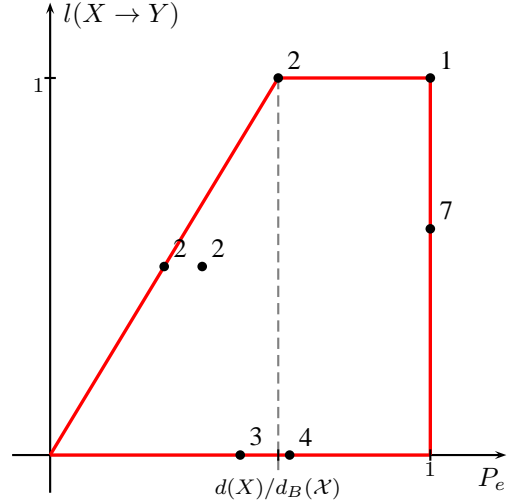

The possible region for a $(\perr,\relLoss{X\to Y})$-pair is depicted in Fig.~\ref{fig:pairregion}. Note that, to our knowledge, this region cannot be restricted further. For example, with reference to Section~\ref{sec:PBFs} there exist systems with $\relLoss{X\to Y}=0$ but with $\perr>0$. Conversely, for a simple projection one will have $\perr=1$ while $\relLoss{X\to Y}<1$. Finally, that $\relLoss{X\to Y}=1$ need not imply $\perr=1$ can be shown by revisiting the center clipper:

\begin{example}[ex:cclipper]{Center Clipper (revisited)}
 Assume that the input probability measure $P_X$ is mixed with an absolutely continuous component supported on $[-c,c]$ ($0<P_X([-c,c])<1$) and a point mass at an arbitrary point $x_0\notin [-c,c]$. According to~\cite{Renyi_InfoDim,Wu_Renyi}, we have $\infodim{X}=P_X([-c,c])$. The output probability measure $P_Y$ has two point masses at 0 and $x_0$ with $P_Y(0)=P_X([-c,c])$ and $P_Y(x_0)=1-P_X([-c,c])$, respectively. Clearly, $\infodim{X|Y=0}=1$ while $\infodim{X|Y=x_0}=0$. Consequently,
\begin{equation}
 \relLoss{X\to Y} = \frac{\infodim{X|Y}}{\infodim{X}} = 1.
\end{equation}
In comparison to that, we have $\perr\leq P_X([-c,c])$, since one can always use the reconstructor $\rec{y}=x_0$ for all $y$.
\end{example}

This is a further example where Proposition~\ref{prop:dimTrans} holds, despite that neither the center clipper is Lipschitz, nor that the requirement of a continuously distributed input RV in Proposition~\ref{prop:RILdimRed} is met.

It is worth mentioning that Proposition~\ref{prop:fanoUBRIL} allows us to prove a converse to lossless analog compression, as it was investigated in~\cite{Wu_Renyi,Wu_PhaseTransition}. To this end, and borrowing the terminology and notation from~\cite{Wu_Renyi}, we \emph{encode} a length-$n$ block $\Xvec^n$ of independent realizations of a real-valued input RV $X$ with information dimension $0<\infodim{X}\leq 1$ via a Lipschitz mapping to the Euclidean space of dimension $\lfloor \mathsf{R} n\rfloor\leq n$. Let $\mathsf{R}(\epsilon)$ be the infimum of $\mathsf{R}$ such that there exists a Lipschitz $g:\mathbb{R}^n\to\mathbb{R}^{\lfloor \mathsf{R} n\rfloor}$ and an arbitrary (measurable) reconstructor $r$ such that $\perr\leq\epsilon$.

\begin{cor}[Converse for Lipschitz Encoders; connection to~\cite{Wu_Renyi},~{\cite[eq.~(26)]{Wu_PhaseTransition}}]\label{cor:converse}
 For a memoryless source with compactly supported marginal distribution $P_X$ and information dimension $0<\infodim{X}\leq 1$, and a Lipschitz encoder function $g$,
\begin{equation}
 \mathsf{R}(\epsilon) \geq \infodim{X}-\epsilon.
\end{equation}
\end{cor}

\begin{IEEEproof}
 Since $\Xvec^n$ is the collection of $n$ real-valued, independent RVs it follows that $\dom{X}^n=\mathbb{R}^n$ and thus $\boxdim{\dom{X}^n}=n$. With Proposition~\ref{prop:fanoUBRIL} we thus obtain
\begin{eqnarray}
n\perr & \geq & \infodim{\Xvec^n} \relLoss{\Xvec^n\to\Yvec^n}\\
&\stackrel{(a)}{=}& \infodim{\Xvec^n}-\infodim{\Yvec^n}\\
&\stackrel{(b)}{=}&  n\infodim{X}-\infodim{\Yvec^n}
\end{eqnarray}
where $(a)$ is due to Proposition~\ref{prop:dimTrans} and $(b)$ is due to the fact that the information dimension of a set of independent RVs is the sum of the individual information dimensions (see, e.g.,~\cite{Cutler_ChainRule} or~\cite[Lem.~3]{Wu_PhD}). Since $\Yvec^n$ is an $\mathbb{R}^{\lfloor \mathsf{R} n\rfloor}$-valued RV, $\infodim{\Yvec^n}\leq \lfloor \mathsf{R} n\rfloor$. Thus,
\begin{equation}
  n\perr \geq n\infodim{X}- \lfloor \mathsf{R} n\rfloor\geq n\infodim{X}-  \mathsf{R} n.
\end{equation}
Dividing by the block length $n$ and rearranging the terms yields
\begin{equation}
 \mathsf{R} \geq \infodim{X}-\perr.
\end{equation}
This completes the proof.
\end{IEEEproof}

While this result -- compared with those presented in~\cite{Wu_Renyi,Wu_PhaseTransition} -- is rather weak, it suggests that our theory has relationships with different topics in information theory, such as compressed sensing. Note further that we need not restrict the reconstructor, since we only consider the case where already the encoder -- the function $g$ -- loses information. The restriction of Lipschitz continuity cannot be dropped, however, since only this class of functions guarantees that Proposition~\ref{prop:dimTrans} holds. In general, as stated in~\cite{Wu_Renyi}, there are non-Lipschitz bijections from $\mathbb{R}^n$ to $\mathbb{R}$.

\subsection{Special Case: 1D-maps and mixed RVs}
We briefly analyze the relative information loss for scenarios similar to the one of Example~\ref{ex:cclipper}: We consider the case where $\dom{X},\dom{Y}\subset\mathbb{R}$, but we drop the restriction that $P_X\ll\mu$. 
%
Instead, we limit ourselves to mixtures of continuous and discrete probability measures, i.e., we assume that $P_X$ has no singular continuous component. Thus,~\cite[pp.~121]{Rudin_Analysis3}
\begin{equation}
 P_X = P_X^{ac} + P_X^d.
\end{equation}
According to~\cite{Renyi_InfoDim,Wu_Renyi} we obtain $\infodim{X}=P_X^{ac}(\dom{X})$. We present

\begin{prop}[Relative Information Loss for Mixed RVs]\label{prop:RILmixed}
 Let $X$ be a mixed RV with a probability measure $P_X = P_X^{ac} + P_X^d$, $0<P_X^{ac}(\dom{X})\leq1$. Let $\{\dom{X}_i\}$ be a finite partition of $\dom{X}\subseteq\mathbb{R}$ into compact sets. Let $g$ be a bounded function such that $g|_{\dom{X}_i}$ is either injective or constant. The relative information loss is given as
\begin{equation}
 \relLoss{X\to Y}=\frac{P_X^{ac}(A)}{P_X^{ac}(\dom{X})}
\end{equation}
where $A$ is the union of sets $\dom{X}_i$ on which $g$ is constant.
\end{prop}

\def \proofRILmixed {0}
\begin{IEEEproof}
\ifthenelse{\proofRILmixed=1}{Since the partition is finite, we can write with~\cite[Thm.~2]{Wu_Renyi}
\begin{equation}
 \infodim{X|Y=y}= \sum_{i} \infodim{X|Y=y,X\in\dom{X}_i} P_{X|Y=y}(\dom{X}_i).
\end{equation}
If $g$ is injective on $\dom{X}_i$, the intersection $\preim{y}\cap\dom{X}_i$ is a single point\footnote{We do not consider the case here that the preimage of $y$ does not intersect $\dom{X}_i$, since in this case $P_{X|Y=y}(\dom{X}_i)=0$.}, thus $\infodim{X|Y=y,X\in\dom{X}_i}=0$. Conversely, if $g$ is constant on $\dom{X}_i$, the preimage is $\dom{X}_i$ itself, so one obtains
\begin{equation}
 \infodim{X|Y=y,X\in\dom{X}_i}=\infodim{X|X\in\dom{X}_i}= \frac{P_X^{ac}(\dom{X}_i)}{P_X(\dom{X}_i)}.
\end{equation}
We thus write
\begin{IEEEeqnarray}{RCL}
 \infodim{X|Y}&=&\int_{\dom{Y}} \sum_{i} \infodim{X|Y=y,X\in\dom{X}_i} P_{X|Y=y}(\dom{X}_i)dP_Y(y)\notag\\\\
&=& \int_\dom{Y} \sum_{i: \dom{X}_i\subseteq A}\frac{P_X^{ac}(\dom{X}_i)}{P_X(\dom{X}_i)}P_{X|Y=y}(\dom{X}_i)dP_Y(y)\\
&\stackrel{(a)}{=}&  \sum_{i: \dom{X}_i\subseteq A}\frac{P_X^{ac}(\dom{X}_i)}{P_X(\dom{X}_i)}\int_\dom{Y}P_{X|Y=y}(\dom{X}_i)dP_Y(y)\\
&\stackrel{(b)}{=}& P_X^{ac}(A)
\end{IEEEeqnarray}
where in $(a)$ we exchanged summation and integration with the help of Fubini's theorem and $(b)$ is due to the fact that the sum runs over exactly the union of sets on which $g$ is constant, $A$. Proposition~\ref{prop:RILDim} completes the proof.
\ifthenelse{\proofRILmixed=1}{}{\endproof}}{See Appendix~\ref{app:proofRILmixed}.}
\end{IEEEproof}

If we compare this result with Corollary~\ref{cor:RILConst}, we see that the former implies the latter for $P_X^{ac}(\dom{X})=1$. Moreover, as we saw in Example~\ref{ex:cclipper}, the relative information loss induced by a function $g$ can increase if the probability measure is not absolutely continuous: In this case, from $P_X([-c,c])$ (where $P_X\ll\mu$) to 1. As we will show next, the relative information loss can also decrease:

\begin{example}[ex:cclipper]{Center Clipper (revisited)} 
Assume that $P_X^{ac}(\dom{X})=0.6$ and $P_X^{ac}([-c,c])=0.3$. The remaining probability mass is a point mass at zero, i.e., $P_X(0)=P_X^d(0)=0.4$. It follows that $\infodim{X}=0.6$ and, from Proposition~\ref{prop:RILmixed}, $\relLoss{X\to Y}=0.5$. We choose a fixed reconstructor $\rec{0}=0$ with a reconstruction error probability $\perr=P_X^{ac}([-c,c])=0.3$. Using Proposition~\ref{prop:fanoUBRIL} we obtain
\begin{equation}
 0.5=\relLoss{X\to Y}\leq \frac{\boxdim{\dom{X}}}{\infodim{X}}\perr = \frac{1}{0.6}0.3=0.5
\end{equation}
which shows that in this case the bound holds with equality.

Consider now the case that the point mass at 0 is split into two point masses at $a,b\in[-c,c]$, where $P_X^d(a)=0.3$ and $P_X^d(b)=0.1$. Using $\rec{0}=a$ the reconstruction error increases to $\perr=P_X([-c,c])-P_X^d(a)=0.4$. Proposition~\ref{prop:fanoUBRIL} now is a strict inequality.
\end{example}

\section{Implications for a System Theory}\label{sec:systemTheory}
In the previous sections we have developed a series of results about the information loss -- absolute or relative -- caused by deterministic systems. In fact, quite many of the basic building blocks of static, i.e., memoryless, systems have been dealt with: Quantizers, the bridge between continuous- and discrete-amplitude systems, have been dealt with in Example~\ref{ex:quantizer}. Cascades of systems allow a simplified analysis employing Propositions~\ref{prop:cascadeLoss} and~\ref{prop:cascadeRelLoss}. Dimensionality reductions of all kinds -- functions which are constant somewhere, omitting coordinates of RVs, etc. -- where a major constituent part of Section~\ref{sec:uncoutPreim}. Finally, the third-order polynomial (Example~\ref{ex:3rd}) is significant in view of Weierstrass' approximation theorem (polynomial functions are dense in the space of continuous functions supported on bounded intervals). Connecting subsystems in parallel and adding the outputs -- a case of dimensionality reduction -- is so common that it deserves separate attention:

\begin{example}{Adding Two RVs}\label{ex:adder}
 We consider two $N$-dimensional input RVs $X_1$ and $X_2$, and assume that the output of the system under consideration is given as
\begin{equation}
 Y=X_1+X_2
\end{equation}
i.e., as the sum of these two RVs.

We start by assuming that $X_1$ and $X_2$ have a joint probability measure $P_{X_1,X_2}\ll\mu^{2N}$. As it can be shown rather easily by transforming $X_1,X_2$ invertibly to $X_1+X_2,X_1$ and dropping the second coordinate, it follows that in this case
\begin{equation}
 \relLoss{X_1,X_2\to Y}=\frac{1}{2}.
\end{equation}

Things may look totally different if the joint probability measure $P_{X_1,X_2}$ is supported on some lower-dimensional submanifold of $\mathbb{R}^{2N}$. Consider, e.g., the case where $X_2=-X_1$, thus $Y\equiv 0$, and $\relLoss{X_1,X_2\to Y}=1$. In contrary to this, assume that both input variables are one-dimensional, and that $X_2=-0.01 X_1^3$. Then, as it turns out,
\begin{equation}
 Y=X_1-0.01 X_1^3 = -0.01 (X_1^3-100X_1)
\end{equation}
which is a piecewise bijective function. As the analysis of Example~\ref{ex:3rd} shows, $\relLoss{X_1,X_2\to Y}=0$ in this case.
\end{example}

We will next apply these results to systems which are larger than the toy examples presented so far. In particular, we focus on an autocorrelation receiver as an example for a system which looses an infinite amount of information, and on an accumulator which will be shown to loose only a finite amount. Yet another example -- the analysis of principle component analysis employing the sample covariance matrix -- can be found in the extended version of~\cite{Geiger_RILPCA_arXiv}. We will not only analyze the information loss in these systems, but also investigate how information \emph{propagates} on the signal flow graph of the corresponding system. This will eventually mark a first step towards a system theory from an information-theoretic point-of-view.

\subsection{Multi-Channel Autocorrelation Receiver}\label{ssec:acr}
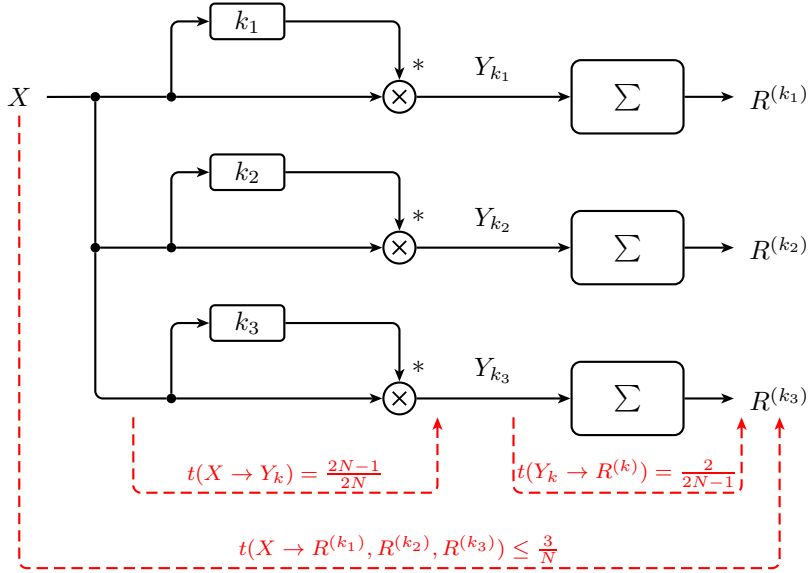
\begin{figure*}
 \centering
\begin{pspicture}[showgrid=false](-8,-3.5)(4,4)
 \psset{style=RoundCorners}
 \pssignal(-7,3){x}{$X$}
 \dotnode(-6,3){dotin1} \dotnode(-6,1){dotin2}
 \ncline{x}{dotin1}\ncline{dotin1}{dotin2}
 \dotnode(-5,3){dotdel1}\dotnode(-5,1){dotdel2}\dotnode(-5,-1){dotdel3}
 \ncline{dotin1}{dotdel1}\ncline{dotin2}{dotdel2}\ncangle[angleA=-90,angleB=180]{dotin2}{dotdel3}
 \psfblock[framesize= 1 0.5](-4,4){del1}{$k_1$}\psfblock[framesize= 1 0.5](-4,2){del2}{$k_2$}\psfblock[framesize= 1 0.5](-4,0){del3}{$k_3$}
 \ncangle[angleA=90,angleB=180,style=Arrow]{dotdel1}{del1}\ncangle[angleA=90,angleB=180,style=Arrow]{dotdel2}{del2}\ncangle[angleA=90,angleB=180,style=Arrow]{dotdel3}{del3}
 \pscircleop[operation=times](-2,3){time1}\pscircleop[operation=times](-2,1){time2}\pscircleop[operation=times](-2,-1){time3}
 \ncangle[angleA=0,angleB=90,style=Arrow]{del1}{time1}\ncangle[angleA=0,angleB=90,style=Arrow]{del2}{time2}\ncangle[angleA=0,angleB=90,style=Arrow]{del3}{time3}
 \ncline[style=Arrow]{dotdel1}{time1}\ncline[style=Arrow]{dotdel2}{time2}\ncline[style=Arrow]{dotdel3}{time3}
 \nput{-300}{time1}{$*$}\nput{-300}{time2}{$*$}\nput{-300}{time3}{$*$}
 \psfblock[framesize= 1.5 1](1,3){sum1}{$\sum$}\psfblock[framesize= 1.5 1](1,1){sum2}{$\sum$}\psfblock[framesize= 1.5 1](1,-1){sum3}{$\sum$}
 \pssignal(3,3){r1}{$R^{(k_1)}$}\pssignal(3,1){r2}{$R^{(k_2)}$}\pssignal(3,-1){r3}{$R^{(k_3)}$}
 \psset{style=Arrow}
 \nclist{ncline}[naput]{time1,sum1 $Y_{k_1}$,r1}\nclist{ncline}[naput]{time2,sum2 $Y_{k_2}$,r2}\nclist{ncline}[naput]{time3,sum3 $Y_{k_3}$,r3}

 \psset{style=Dash,linecolor=red}
 \psline(-5.5,-1.25)(-5.5,-2.25)(-1.5,-2.25)(-1.5,-1.25)\rput[c]{0}(-3.5,-2){\footnotesize\textcolor{red}{$\relTrans{X\to Y_k}=\frac{2N-1}{2N}$}}
 \psline(-0.5,-1.25)(-0.5,-2.25)(2.5,-2.25)(2.5,-1.25)\rput[c]{0}(1,-2){\footnotesize\textcolor{red}{$\relTrans{Y_k\to R^{(k)}}=\frac{2}{2N-1}$}}
 \psline(-7,2.75)(-7,-3.25)(3,-3.25)(3,-1.25)\rput[c]{0}(-2,-3){\footnotesize\textcolor{red}{$\relTrans{X\to R^{(k_1)},R^{(k_2)},R^{(k_3)}}\leq\frac{3}{N}$}}
\end{pspicture}
\caption{Discrete-time model of the multi-channel autocorrelation receiver: The elementary mathematical operations depicted are the complex conjugation ($*$), the summation of vector elements ($\sum$), and the circular shift (blocks with $k_i$). The information flow is illustrated using red arrows, labeled according to the relative information transfer.} 
\label{fig:acr}
\end{figure*}

The multi-channel autocorrelation receiver (MC-AcR) was introduced in~\cite{Witrisal_AcR} and analyzed in~\cite{Meissner_AcR,Pedross_AcR} as a non-coherent receiver architecture for ultrawide band communications. In this receiver the decision metric is formed by evaluating the autocorrelation function of the input signal for multiple time lags (see Fig.~\ref{fig:acr}).

To simplify the analysis, we assume that the input signal is a discrete-time, complex-valued $N$-periodic signal superimposed with independent and identically distributed complex-valued noise. The complete analysis can be based on $N$ consecutive values of the input, which we will denote with $\X{1}$ through $\X{N}$ ($X$ is the collection of these RVs). The real and imaginary parts of $\X{i}$ will be denoted as $\Re\X{i}$ and $\Im\X{i}$, respectively. We may assume that $P_X\ll\mu^{2N}$, where $P_X$ is compactly supported. We consider only three time lags $k_1,k_2,k_3\in\{1,\dots,N-1\}$.

By assuming periodicity of the input signal, we can replace the linear autocorrelation by the circular autocorrelation (e.g.,~\cite[pp.~655]{Oppenheim_Discrete3})
\begin{equation}
 R^{(k)} = \sum_{n=0}^{N-1} \X{n}X^{*(n+k)} = \sum_{n=0}^{N-1} \Y{n}_k
\end{equation}
where $*$ denotes complex conjugation, and where $k$ assumes one value of the set $\{k_1,k_2,k_3\}$. Note further that the circular autocorrelation here is implicit, since $\X{n}=\X{n+N}$.

We start by noting that for $k=0$, $\relLoss{X\to Y_0}=\frac{1}{2}$, since then
\begin{equation}
  \Y{n}_0=\X{n}X^{*(n)}=|\X{n}|^2
\end{equation}
which shows that $Y_0$ is real and thus $P_{Y_0}\ll\mu^N$. Since $R^{(0)}$ is the sum of the components of $Y_0$, it is again real and we get
\begin{equation}
 \relTrans{X\to R^{(0)}}=\frac{1}{2N}.
\end{equation}

\begin{figure}
 \centering
\begin{pspicture}[showgrid=false](0,-5)(8,-1)
 \psset{style=RoundCorners,style=Arrow}
  \pssignal(0,-3){xh}{$X$}
 \psfblock[framesize= 1 0.75](1.5,-3){log}{$\log(\cdot)$}
 \dotnode(3,-3){dotlog}
 \psfblock[framesize= 1 0.5](4,-2){logdel}{$k_1$}
 \pscircleop(4.8,-3){logtime}\nput{-300}{logtime}{$*$}
 \psfblock[framesize= 1 0.75](6.5,-3){e}{$\e{(\cdot)}$}
 \pssignal(8,-3){y}{$Y_{k_1}$}
 \nclist{ncline}{xh,log,logtime,e,y}
 \rput{0}(2.5,-2.75){$\tilde{X}$}\rput{0}(5.5,-2.75){$\tilde{Y}_{k_1}$}
 \ncangle[angleA=90,angleB=180]{dotlog}{logdel}\ncangle[angleA=0,angleB=90]{logdel}{logtime}
 \psset{style=Dash,linecolor=red}
 \psline(0.5,-2.5)(0.5,-1.5)(2.5,-1.5)(2.5,-2.5)\rput[c]{0}(1.5,-1.2){\footnotesize\textcolor{red}{$\loss{X\to \cdot}=0$}}
 \psline(5.5,-2.5)(5.5,-1.5)(7.5,-1.5)(7.5,-2.5)\rput[c]{0}(6.5,-1.2){\footnotesize\textcolor{red}{$\loss{\cdot\to Y_k}=0$}}
 \psline(2.5,-3.2)(2.5,-4.2)(5.5,-4.2)(5.5,-3.2)\rput[c]{0}(4,-3.9){\footnotesize\textcolor{red}{$\relLoss{\cdot\to\cdot}=\frac{1}{2}$}}
\end{pspicture}
\caption{Equivalent model for multiplying the two branches in Fig.~\ref{fig:acr}.}
\label{fig:logstuff}
\end{figure}
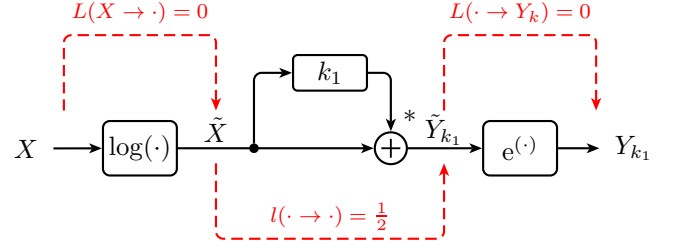

\newcommand{\Xt}[1]{\tilde{X}^{(#1)}}
For non-zero $k$ note that (see Fig.~\ref{fig:logstuff})
\begin{multline}
 \X{n}X^{*(n+k)} = \e{\log\X{n}X^{*(n+k)}}\\
= \e{\log\X{n}+\log X^{*(n+k)}}=\e{\log\X{n}+(\log\X{n+k})^*}\\
= \e{\Xt{n}+\tilde{X}^{*(n+k)}} = \e{\tilde{Y}^{(n)}_k} = \Y{n}_k.
\end{multline}
In other words, we can write the multiplication as an addition (logarithm and exponential function are invertible and, thus, information lossless).

Letting $\tilde{X}_k$ denote the vector of elements indexed by $\Xt{n+k}$, we get
\begin{equation}
  \tilde{X}_k = \mathbf{C}_k \tilde{X}
\end{equation}
where $\mathbf{C}_k$ is a circulant permutation matrix. Thus, $\tilde{Y}_k=\tilde{X}+\overline{\tilde{X}_k}$ and
\begin{IEEEeqnarray}{RCL}
 \Re \tilde{Y}_k &=& (\eye+\mathbf{C}_k) \Re \tilde{X}\\
\Im \tilde{Y}_k &=& (\eye-\mathbf{C}_k) \Im \tilde{X}
\end{IEEEeqnarray}
where $\eye$ is the $N\times N$ identity matrix. Since $\eye+\mathbf{C}_k$ is invertible, we have $P_{\Re \tilde{Y}_k}\ll\mu^N$. In contrary to that, the rank of $\eye-\mathbf{C}_k$ is $N-1$
and thus $\infodim{\Im \tilde{Y}_k}=N-1$. It follows that
\begin{multline}
 \relTrans{\Re \tilde{X},\Im \tilde{X} \to \Re \tilde{Y}_k,\Im \tilde{Y}_k} = \relTrans{\tilde{X}\to \tilde{Y}_k}\\= \relTrans{X\to Y_k} = \frac{2N-1}{2N}
\end{multline}
and $\infodim{Y_k}=2N-1$.

Clearly, for $k\neq 0$ the autocorrelation will be a complex number a.s., thus $\infodim{R^{(k)}}=2$. Since the summation is a Lipschitz function, we obtain
\begin{equation}
 \relTrans{Y_k\to R^{(k)}} = \frac{2}{2N-1}
\end{equation}
and by the result about the cascades,
\begin{equation}
 \relTrans{X\to R^{(k)}} = \frac{1}{N}.
\end{equation}
Finally, applying Proposition~\ref{prop:UBRelTrans},
\begin{equation}
 \relTrans{X\to R^{(k_1)},R^{(k_2)},R^{(k_3)}} \leq \frac{3}{N}.
\end{equation}
Note that this analysis would imply that if all values of the autocorrelation function would be evaluated, the relative information transfer would increase to
\begin{equation}
 \relTrans{X\to R} \leq \frac{2N-1}{2N}.
\end{equation}
However, knowing that the autocorrelation function of a complex, periodic sequence is Hermitian and periodic with the same period, it follows that $P_R\ll\mu^N$, and thus $\relTrans{X\to R}=\frac{1}{2}$. The bound is thus obviously not tight in this case. Note further that, applying the same bound to the relative information transfer from $X$ to $Y_{k_1},Y_{k_2},Y_{k_3}$ would yield a number greater than one. This is simply due to the fact that the three output vectors have a lot of information in common, prohibiting simply adding their information dimensions.

\begin{figure}
 \centering
\begin{pspicture}[showgrid=false](0,-5)(8.5,-1)
 \psset{style=RoundCorners,style=Arrow}
  \pssignal(0,-3){xh}{$X$}
 \psfblock[framesize= 1 0.75](1.25,-3){log}{$\mathsf{DFT}$}
 \psfblock[framesize= 1 0.75](3,-3){logdel}{$|\cdot|^2$}
 \psfblock[framesize= 1 0.75](4.75,-3){e}{$\mathsf{IDFT}$}
 \psfblock[framesize= 1 0.75](6.5,-3){pi}{$\Pi$}
 \pssignal(8.5,-3){y}{$\begin{array}{ccc}
                        R^{(k_1)}\\R^{(k_2)}\\R^{(k_3)}
                       \end{array}$}
 \nclist{ncline}[naput]{xh,log,logdel $F_X$,e $F_R$,pi $R$,y}
 \ncangle[angleA=90,angleB=180]{dotlog}{logdel}\ncangle[angleA=0,angleB=90]{logdel}{logtime}
 \psset{style=Dash,linecolor=red}
 \psline(0,-2.5)(0,-1.5)(2.125,-1.5)(2.125,-2.25)\rput[c]{0}(1.06,-1.2){\footnotesize\textcolor{red}{$\loss{X\to F_X}=0$}}
 \psline(3.875,-2.25)(3.875,-1.5)(5.625,-1.5)(5.625,-2.25)\rput[c]{0}(4.75,-1.2){\footnotesize\textcolor{red}{$\loss{F_R\to R}=0$}}
 \psline(2.125,-3.5)(2.125,-4.25)(3.875,-4.25)(3.875,-3.5)\rput[c]{0}(3,-4.5){\footnotesize\textcolor{red}{$\relTrans{F_X\to F_R}=\frac{1}{2}$}}
 \psline(5.625,-3.5)(5.625,-4.25)(8.5,-4.25)(8.5,-3.5)\rput[c]{0}(7,-4.5){\footnotesize\textcolor{red}{$\relTrans{R\to \cdot}\leq\frac{6}{N}$}}
\end{pspicture}
\caption{Equivalent model of the MC-AcR of Fig.~\ref{fig:acr}, using the DFT to compute the circular autocorrelation. $\mathsf{IDFT}$ denotes the inverse DFT and $\Pi$ a projection onto a subset of coordinates. The information flow is indicated by red arrows labeled according to the relative information transfer.}
\label{fig:acrDFT}
\end{figure}
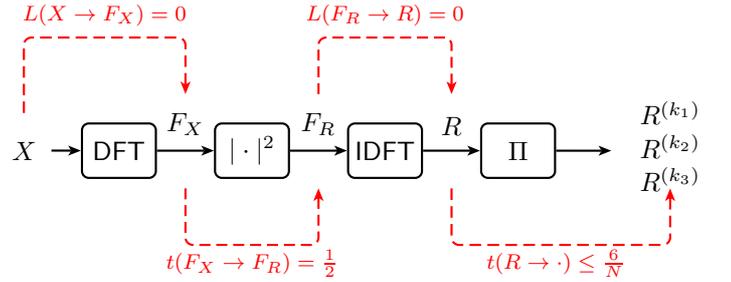

A slightly different picture is revealed if we look at an equivalent signal model, where the circular autocorrelation is computed via the discrete Fourier transform (DFT, cf.~Fig.~\ref{fig:acrDFT}): Letting $\Wvec$ denote the DFT matrix, we obtain the DFT of $X$ as $F_X=\Wvec X$. Doing a little algebra, the DFT of the autocorrelation function is obtained as
\begin{equation}
 F_R = |F_X|^2.
\end{equation}
Since the DFT is an invertible transform ($\Wvec$ is a unitary matrix), one has $\infodim{F_X}=\infodim{X}=2N$. The squaring of the magnitude can be written as a cascade of a coordinate transform (from Cartesian to polar coordinates; invertible), a dimensionality reduction (the phase information is dropped), and a squaring function (invertible, since the magnitude is a non-negative quantity). It follows that
\begin{equation}
 \relLoss{X\to R}=\relLoss{F_X\to F_R}=\frac{1}{2}.
\end{equation}
since the inverse DFT is again invertible.

Since $F_R$ is a real RV and thus $P_{F_R}\ll\mu^N$, it clearly follows that also $P_R\ll\mu^N$, despite the fact that $R$ is a vector of $N$ complex numbers. The probability measure of this RV is concentrated on an $N$-dimensional submanifold of $\mathbb{R}^{2N}$ defined by the periodicity and Hermitian symmetry of $R$. Choosing three coordinates of $R$ as the final output of the system amounts to upper bounding the information dimension of the output by
\begin{equation}
 \infodim{R^{(k_1)},R^{(k_2)},R^{(k_3)}}\leq 6.
\end{equation}
Thus,
\begin{equation}
 \relTrans{X\to R^{(k_1)},R^{(k_2)},R^{(k_3)}} \leq \frac{3}{N}
\end{equation}
where equality is achieved if, e.g., all time lags are distinct and smaller than $\frac{N}{2}$. The information flow for this example -- computed from the relative information loss and information transfer -- is also depicted in Figs.~\ref{fig:acr} and~\ref{fig:acrDFT}.

\subsection{Accumulator}\label{ssec:accumulator}
\newcommand{\Si}[1]{S^{(#1)}}

\begin{figure}[t]
 \centering  
\begin{pspicture}[showgrid=false](-1,0)(7,3)
  \psset{style=RoundCorners,style=Arrow}
  \pssignal(-1,2){px}{$p_X$}\pssignal(7,2){ps}{$p_{\Si{i}}$}
  \pssignal(1,2){x}{$\X{i}$}
  \ncline{px}{x}
  \pscircleop(2.75,2){oplus1}
  \psfblock[framesize=1 0.75](4,0.5){oplus}{$z^{-1}$}
  \ncline[style=Arrow]{x}{oplus1}
  \dotnode(5.25,2){dot1}
  \ncline{oplus1}{dot1}
  \ncangle[angleA=180,angleB=-90,style=Arrow]{oplus}{oplus1}\aput*{0}{$\Si{i-1}$}
  \ncangle[angleA=-90,angleB=0,style=Arrow]{dot1}{oplus}\aput*{0}{$\Si{i}$}
  \ncline{dot1}{ps}
  \fnode[style=Dash,linecolor=red,framesize=6 2.6](3,1.1){box}
  \nput{90}{box}{\textcolor{red}{$g$}}
\end{pspicture}
\caption{Accumulating a sequence of independent, identically distributed RVs $\X{i}$}
\label{fig:accu}
\end{figure}
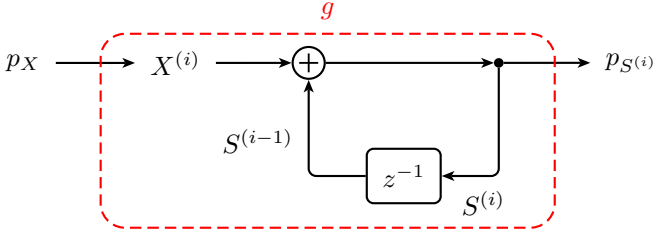

As a further example, consider the system depicted in Fig.~\ref{fig:accu}. We are interested in how much information we loose about a probabiliy measure of an RV $X$ if we observe the probability measure of a sum of iid copies of this RV. For simplicity, we assume that the probability measure of $X$ is supported on a finite field $\dom{X}=\{0,\dots,N-1\}$, where we assume $N$ is even. 

The input to the system is thus the probability mass function (PMF) $p_X$. Since its elements must sum to unity, the vector $p_X$ can be chosen from the $(N-1)$-simplex in $\mathbb{R}^N$; we assume $P_{p_X}\ll\mu^{N-1}$. The output of the system at a certain discrete time index $i$ is the PMF $p_{\Si{i}}$ of the sum $\Si{i}$ of $i$ iid copies of $X$:
\begin{equation}
 \Si{i} = \bigoplus_{k=1}^i \X{k}
\end{equation}
where $\bigoplus$ denotes modulo-addition.

Given the PMF $p_{\Si{i}}$ of the output $\Si{i}$ at some time $i$ (e.g., by computing the histogram of multiple realizations of this system), how much information do we lose about the PMF of the input $X$? Mathematically, we are interested in the following quantity\footnote{Note that $\loss{p_X\to p_{\Si{i}}}$ is not to be confused with $\ent{\Si{1}|\Si{i}}$, a quantity measuring the information loss about the initial \emph{state} of a Markov chain.}:
\begin{equation}
 \loss{p_X\to p_{\Si{i}}}
\end{equation}

From the theory of Markov chains we know that $p_{\Si{i}}$ converges to a uniform distribution on $\dom{X}$. To see this, note that the transition from $\Si{i}$ to $\Si{i+1}$ can be modeled as a cyclic random walk; the transition matrix of the corresponding Markov chain is a positive circulant matrix built\footnote{In this particular case it can be shown that the first row of the transition matrix consists of the elements of $p_X$, while all other rows are obtained by circularly shifting the first one.} from $p_X$. As a consequence of the Perron-Frobenius theorem (e.g.,~\cite[Thm.~15-8]{Papoulis_Probability}) there exists a unique stationary distribution which, for a doubly stochastic matrix as in this case, equals the uniform distribution on $\dom{X}$ (cf.~\cite[Thm.~4.1.7]{Kemeny_FMC}).

To attack the problem, we note that the PMF of the sum of independent RVs is given as the convolution of the PMFs of the summands. In case of the modulo sum, the circular convolution needs to be applied instead, as can be shown by computing one Markov step. Using the DFT again, we can write the circular convolution as a multiplication; in particular (see, e.g.,~\cite[Sec.~8.6.5]{Oppenheim_Discrete3})
\begin{equation}
 (p_X * p_{\Si{i}}) \longleftrightarrow F_{p_X}F_{p_{\Si{i}}}
\end{equation}
where $F_{p_X}=\Wvec p_X$. Iterating the system a few time steps and repeating this analysis yields
\begin{equation}
 p_{\Si{i}} = \Wvec^{-1} F_{p_X}^i
\end{equation}
where the $i$-th power is taken element-wise. Since neither DFT nor inverse DFT lose information, we only have to consider the information lost in taken $F_{p_X}$ to the $i$-th power.

We now employ a few properties of the DFT (see, e.g.,~\cite[Sec.~8.6.4]{Oppenheim_Discrete3}): Since $p_X$ is a real vector, $F_{p_X}$ will be Hermitian (circularly) symmetric; moreover, if we use indices from $0$ to $N-1$, we have $F_{p_X}^{(0)}=1$ and $F_{p_X}^{(\frac{N}{2})}\in\mathbb{R}$. 

Taking the $i$-th power of a real number $\beta$ looses at most one bit of information for even $i$ and nothing for odd $i$; thus
\begin{equation}
 \loss{\beta\to \beta^i} \le \cos^2\left(\frac{i\pi}{2}\right). \label{eq:realpower}
\end{equation}

Taking the power of a complex number $\alpha$ corresponds to taking the power of its magnitude (which can be inverted by taking the corresponding root) and multiplying its phase. Only the latter is a non-injective operation, since the $i$-th root of a complex number yields $i$ different solutions for the phase. Thus, invoking Proposition~\ref{prop:UBLoss}
\begin{equation}
 \loss{\alpha \to\alpha^i} \leq \log i. \label{eq:complexpower}
\end{equation}

Applying~\eqref{eq:complexpower} to index $\frac{N}{2}$ and~\eqref{eq:complexpower} to the indices $\{1,\dots,\frac{N}{2}-1\}$ reveals that
\begin{multline}
 \loss{p_X\to p_{\Si{i}}} = \loss{F_{p_X}\to F_{p_{\Si{i}}}} \\\leq \left(\frac{N}{2}-1\right) \log i+\cos^2\left(\frac{i\pi}{2}\right).
\end{multline}
Thus, the information loss increases linearly with the number of components, but sublinearly with time. Moreover, since for $i\to\infty$, $p_{\Si{i}}$ converges to a uniform distribution, it is plausible that $\loss{p_X\to p_{\Si{i}}}\to \infty$ (at least for $N>2$). Intuitively, the earlier one observes the output of such a system, the more information about the unknown PMF $p_X$ can be retrieved. 


\subsection{Metrics of Information Flow -- Do We Need Them All?}
So far, for our system theory we have introduced absolute and relative information loss, as well as relative information transfer. Together with mutual information this makes four different metrics which can be used to characterize a deterministic input-output system. While clearly relative information loss and relative information transfer are equivalent, the previous examples showed that one cannot simple omit the other measures of information flow without losing flexibility in describing systems. Assuming that only finite measures are meaningful, a quantizer, e.g., requires mutual information, whereas a rectifier would need information loss. Center clippers or other systems for which both absolute measures are infinite, benefit from relative measures only (be it either relative information transfer or relative information loss). We summarize notable examples from this work together with their adequate information measures in Table~\ref{tab:comparison}.

\begin{table}[t]
 \centering
 \caption{Comparison of different information flow measures. The input is assumed to have positive information dimension. The letter $c$ indicates that the measure evaluates to a finite constant.}
\label{tab:comparison}
\begin{tabular}{l|cccc}
 System & $\mutinf{X;Y}$ & $\loss{X\to Y}$ &$\relTrans{X\to Y}$ & $\relLoss{X\to Y}$\\
\hline
Quantizer & $c$ & $\infty$ & 0 & 1\\
Rectifier & $\infty$ & $c$ & 1 & 0\\
Center Clipper & $\infty$ &$\infty$ & $c$ & $1-c$\\
Example~\ref{ex:infloss} & $\infty$ &$\infty$ & $1$ & $0$\\
MC-AcR & $\infty$ &$\infty$ & $c$ & $1-c$\\
Accumulator & $\infty$ &$c$ & 1 & 0
\end{tabular}
\end{table}

\section{Open Issues and Outlook}\label{sec:outlook}
While this work may mark a step towards a system theory from an information-theoretic point-of-view, it is but a small step: In energetic terms, it would ``just'' tell us the difference between the signal variances at the input and the output of the system -- a very simple, memoryless system. We do not yet know anything about the information-theoretic analog of power spectral densities (e.g., entropy rates), about systems with memory, or about the analog of more specific energy measures like the mean-squared reconstruction error assuming a signal model with noise (information loss ``relevant'' in view of a signal model). Moreover, we assumed that the input signal is sufficiently well-behaved in some probabilistic sense. Future work should mainly deal with extending the scope of our system theory.

The first issue which will be addressed is the fact that at present we are just measuring information ``as is''; every bit of input information is weighted equally, and losing a sign bit amounts to the same information loss as losing the least significant bit in a binary expansion of a (discrete) RV. This fact leads to the apparent counter-intuitivity of some of our results: To give an example from~\cite{Geiger_RILPCA_arXiv}, the principle component analysis (PCA) applied prior to dimensionality reduction does not decrease the relative information loss\footnote{This opposes the intuition that preserving the subspace with the largest variance should also preserve most of the information: For example, the Wikipedia article~\cite{Wikipedia_PCA} states, among other things, that ``[...] PCA can supply the user with a lower-dimensional picture, a 'shadow' of this object when viewed from its (in some sense) most informative viewpoint'' and that the variances of the dropped coordinates ``tend to be small and may be dropped with minimal loss of information''.}; this loss is always fully determined by the information dimension of the input and the information dimension of the output (cf.~Corollary~\ref{cor:RILProj}). Contrary to that, the literature employs information theory to prove the optimality of PCA in certain cases~\cite{Linsker_Infomax,Plumbley_TN,Deco_ITNN} -- but see also~\cite{Rao_PCA} for a recent work presenting conditions for the PCA depending on the spectrum of eigenvalues for a certain signal-noise model. To build a bridge between our theory of information loss and the results in the literature, the notion of \emph{relevance} has to be brought into game, allowing us to place unequal weights to different portions of the information available at the input. In energetic terms: Instead of just comparing variances -- which is necessary sometimes! -- we are now interested, e.g., in a mean-squared reconstruction error w.r.t. some relevant portion of the input signal. We actually proposed the corresponding notion of \emph{relevant} information loss in~\cite{Geiger_Relevant_arXiv}, where we showed its applicability in signal processing and machine learning and, among other things, re-established the optimality of PCA given a specific signal model. We furthermore showed that this notion of relevant information loss is fully compatible with what we present in this paper.

Going from variances to power spectral densities, or, from information loss to information loss rates, will represent the next step: If the input to our memoryless system is not a sequence of independent RVs but a discrete-time stationary stochastic process, how much information do we lose per unit time? Following~\cite{Watanabe_InfoLoss}, the information loss rate should be upper bounded by the information loss (assuming the marginal distribution of the process as the distribution of the input RV). Aside from this, little is known about this scenario, and we hope to bring some light into this issue in the future. Of particular interest would be the reconstruction of nonlinearly distorted sequences, extending Sections~\ref{ssec:reconLoss} and~\ref{ssec:reconRIL}.

The next, bigger step is from memoryless to dynamical input-output systems: A particularly simple subclass of these are linear filters, which were already analyzed by Shannon\footnote{To be specific, in his work~\cite{Shannon_TheoryOfComm} he analyzed the \emph{entropy loss} in linear filters.}. In the discrete-time version taken from~\cite[pp.~663]{Papoulis_Probability} one gets the differential entropy rate of the output process $\Yvec$ by adding a system-dependent term to the differential entropy rate of the input process $\Xvec$, or
\begin{equation}
 \derate{Y} = \derate{X} +\frac{1}{2\pi} \int_{-\pi}^\pi \log |H(\e{\jmath\theta})|d\theta 
\end{equation}
where $H(\e{\jmath\theta})$ is the frequency response of the filter. The fact that the latter term is independent of the process statistics shows that it is only related to the change of variables, and not to information loss. In that sense, linear filters do not perform information processing where the change of information measures should obviously depend on the input signal statistics.

Finally, the class of nonlinear dynamical systems is significantly more difficult. We were able to present some results for discrete alphabets in~\cite{Geiger_NLDyn1starXiv}. For more general process alphabets we can only hope to obtain results for special subclasses, e.g., Volterra systems or affine input systems. For example, Wiener and Hammerstein systems, which are cascades of linear filters and static nonlinear functions, can completely be dealt with by generalizing our present work to stochastic processes.

Finally, many other aspects are worth investigating: The connection between information loss and entropy production in iterated function systems~\cite{Ruelle_EntropyProduction} and to heat dissipation (Landauer's principle~\cite{Landauer_Principle}) could be of interest. Yet another interesting point is the connection between energetic and information-theoretic measures, as it exists for the Gaussian distribution.

\section{Conclusion}
We presented an information-theoretic way to characterize the behavior of deterministic, memoryless input-output systems. In particular, we defined an absolute and a relative measure for the information loss occurring in the system due to its potential non-injectivity. Since the absolute loss can be finite for a subclass of systems despite a continuous-valued input, we were able to derive Fano-type inequalitites between the information loss and the probability of a reconstruction error.

The relative measure of information loss, introduced in this work to capture systems in which an infinite amount of information is lost (and, possibly, preserved), was shown to be related to R\'{e}nyi's information dimension and to present a lower bound on the reconstruction error probability. With the help of an example we showed that this bound can be tight and that, even in cases where an infinite amount of information is lost, the probability of a reconstruction error need not be unity.

While our theoretical results were developed mainly in view of a system theory, we believe that some of them may be of relevance also for analog compression, reconstruction of nonlinearly distorted signals, chaotic iterated function systems, and the theory of Perron-Frobenius operators.

\appendices
\section{Proof of Proposition~\ref{prop:dimTrans}}\label{app:proofLip}
For the proof we need the following Lemma:
\begin{lem}\label{lem:Lip}
 Let $X$ and $Y$ be the input and output of a Lipschitz function $g{:}\ \dom{X}\to\dom{Y}$, $\dom{X}\subseteq\mathbb{R}^N$, $\dom{Y}\subseteq\mathbb{R}^M$. Then,
\begin{equation}
 \limn \frac{\ent{\hat{Y}_n|\hat{X}_n}}{n} = 0.
\end{equation}
\end{lem}

\begin{IEEEproof}
We provide the proof by showing that $\ent{\hat{Y}_n|\hat{X}_n=\hat{x}_k}$ is finite for all $n$ and for all $\hat{x}_k$. From this then immediately follows that
\begin{multline}
 \limn \frac{\ent{\hat{Y}_n|\hat{X}_n}}{n}\\ = \limn \frac{1}{n}\sum_{k} \ent{\hat{Y}_n|\hat{X}_n=\hat{x}_k} P_X(\hat{\dom{X}}_k^{(n)})=0.
\end{multline}

To this end, note that the conditional probability measure $P_{X|\hat{X}_n=\hat{x}_k}$ is supported on $\hat{\dom{X}}_k^{(n)}$, and that, thus, $P_{Y|\hat{X}_n=\hat{x}_k}$ is supported on $g(\hat{\dom{X}}_k^{(n)})$. Since $g$ is Lipschitz, there exists a constant $\lambda$ such that for all $a,b\in\hat{\dom{X}}_k^{(n)}$
\begin{equation}
 |g(a)-g(b)|\leq\lambda|a-b|.
\end{equation}
Choose $a$ and $b$ such that the term on the left is maximized, i.e., 
\begin{equation}
 \sup_{a,b}|g(a)-g(b)| = |g(a^\circ)-g(b^\circ)| \leq \lambda|a^\circ-b^\circ|\leq \lambda\sup_{a,b}|a-b|
\end{equation}
or, in other words,
\begin{equation}
 \diam{g(\hat{\dom{X}}_k^{(n)})}\leq\lambda\diam{\hat{\dom{X}}_k^{(n)}}\leq \frac{\lambda\sqrt{N}}{2^n}.
\end{equation}
The latter inequality follows since $\hat{\dom{X}}_k^{(n)}$ is inside an $N$-dimensional hypercube of side length $\frac{1}{2^n}$ (one may have equality in the last statement if the hull of $\hat{\dom{X}}_k^{(n)}$ and the boundary of $\dom{X}$ are disjoint).

Now note that the support of $P_{Y|\hat{X}_n=\hat{x}_k}$ can be covered by an $M$-dimensional hypercube of side length $\diam{g(\hat{\dom{X}}_k^{(n)})}$, which can again be covered by
\begin{equation}
 \left\lceil2^n\diam{g(\hat{\dom{X}}_k^{(n)})}+1\right\rceil^M
\end{equation}
$M$-dimensional hypercubes of side length $\frac{1}{2^n}$. By the maximum entropy property of the uniform distribution we get
\begin{multline}
 \ent{\hat{Y}_n|\hat{X}_n=\hat{x}_k}\leq \log \left\lceil2^n\diam{g(\hat{\dom{X}}_k^{(n)})}+1\right\rceil^M\\
\leq M\log \left\lceil2^n\frac{\lambda\sqrt{N}}{2^n}+1 \right\rceil = M\log \left\lceil\lambda\sqrt{N}+1 \right\rceil< \infty.
\end{multline}
This completes the proof.
\end{IEEEproof}

We now turn to the
\begin{IEEEproof}[Proof of Proposition~\ref{prop:dimTrans}]
 By noticing that $\hat{Y}_n$ is a function of $Y$ and using the chain rule of mutual information we expand the term in Definition~\ref{def:relTrans} as
\begin{align}
 \relTrans{X\to Y} &= \limn \frac{\mutinf{\hat{X}_n;\hat{Y}_n}+\mutinf{\hat{X}_n;Y|\hat{Y}_n}}{\ent{\hat{X}_n}}\\
&= \limn \frac{\mutinf{\hat{X}_n;\hat{Y}_n}/n+\mutinf{\hat{X}_n;Y|\hat{Y}_n}/n}{\ent{\hat{X}_n}/n}
\\&=\limn \frac{\ent{\hat{Y}_n}/n+\mutinf{\hat{X}_n;Y|\hat{Y}_n}/n}{\ent{\hat{X}_n}/n}\label{eq:proofTrans}
\end{align}
due to Lemma~\ref{lem:Lip}. 

We now note that, by Kolmogorov's formula,
\begin{IEEEeqnarray}{RCL}
 \mutinf{\hat{X}_n;Y|\hat{Y}_n} &=& \mutinf{\hat{X}_n,\hat{Y}_n;Y}-\mutinf{\hat{Y}_n;Y}\\
 \mutinf{X;Y|Y} &=& \mutinf{X,Y;Y}-\mutinf{Y;Y}\stackrel{(a)}{=}0
\end{IEEEeqnarray}
where $(a)$ is due to the fact that $X-Y-Y$ is a Markov chain (see, e.g.,~\cite[pp.~43]{Pinsker_InfoEngl}). With~\cite[Lem.~7.22]{Gray_Entropy} and the way $\hat{X}_n$ is constructed from $X$ (cf.~\cite[Sec.~III.D]{Wu_Renyi}) we find that $\limn \mutinf{\hat{X}_n,\hat{Y}_n;Y}=\mutinf{X,Y;Y}$ and $\limn\mutinf{\hat{Y}_n;Y}=\mutinf{Y;Y}$, and that, thus\footnote{The authors thank Siu-Wai Ho for suggesting this proof method.}
\begin{equation}
 \limn  \mutinf{\hat{X}_n;Y|\hat{Y}_n}=0.
\end{equation}
By employing Definition~\ref{def:infodim} the proof is completed.
\end{IEEEproof}

\ifthenelse{\proofILDIFFhere=1}{}{
\section{Proof of Proposition~\ref{prop:lossPBFdiffEnt}}\label{app:proofILDIFF}
}

\ifthenelse{\proofWhere=1}{}{
\section{Proof of Proposition~\ref{prop:partitionLoss}}\label{app:proofW}
}

\ifthenelse{\proofUBhere=1}{}{
\section{Proof of Proposition~\ref{prop:UBLoss}}\label{app:proofUB}
}

\ifthenelse{\proofMAPFanohere=1}{}{
\section{Proof of Proposition~\ref{prop:fanoLBLoss}}\label{app:proofMAPFano}
 The proof follows closely the proof of Fano's inequality~\cite[pp.~38]{Cover_Information2}, where one starts with noticing that
\begin{equation}
 \ent{X|Y}=\ent{E|Y}+\ent{X|E,Y}.
\end{equation}
The first term, $\ent{E|Y}$ can of course be upper bounded by $\ent{E}=\binent{\perr}$, as in Fano's inequality. However, also
\begin{multline}
 \ent{E|Y} = \int_{\dom{Y}} \binent{\perr(y)} dP_Y(y)\\
=\int_{\dom{Y}\setminus\dom{Y}_b} \binent{\perr(y)} dP_Y(y)\\
\leq \int_{\dom{Y}\setminus\dom{Y}_b} dP_Y(y) = 1-P_b
\end{multline}
since $\binent{\perr(y)}=\perr(y)=0$ if $y\in\dom{Y}_b$ and since $\binent{\perr(y)}\leq 1$ otherwise. Thus,
\begin{equation}
 \ent{E|Y}\leq\min\{\binent{\perr},1-P_b\}.
\end{equation}

For the second part note that $\ent{X|E=0,Y=y}=0$, so we obtain
\begin{equation}
 \ent{X|E,Y} = \int_\dom{Y}\ent{X|E=1,Y=y}\perr(y)dP_Y(y).
\end{equation}
Upper bounding the entropy by $\log\left(\card{\preim{y}}-1\right)$ we get
\begin{align}
 &\ent{X|E,Y} \leq
\perr\int_\dom{Y}\log\left(\card{\preim{y}}-1\right)\frac{\perr(y)}{\perr}dP_Y(y)\\
&\stackrel{(a)}{\leq} \perr\log\left(\int_\dom{Y}\left(\card{\preim{y}}-1\right)\frac{\perr(y)}{\perr}dP_Y(y)\right)\\
&\stackrel{(b)}{\leq} \perr\log\left(\int_\dom{Y}\left(\card{\preim{y}}-1\right)dP_Y(y)\right)+\perr\log\frac{1}{\perr}\label{eq:dummy1}
\end{align}
where $(a)$ is Jensen's inequality ($\perr(y)/\perr$ acts as a PDF) and $(b)$ holds since $\perr(y)\leq 1$ and due to splitting the logarithm. 
This completes the proof.
\ifthenelse{\proofMAPFanohere=1}{}{\endproof}}

\section{Proof of Proposition~\ref{prop:Suboptimal}}\label{app:proofFano}
By construction, $\recsub{y}=x$ whenever $x\in\dom{X}_k\cup\dom{X}_b$, and conversely, $\recsub{y}\neq x$ whenever $x\notin\dom{X}_k\cup\dom{X}_b$. This yields $\perh=1-P_X(\dom{X}_k\cup\dom{X}_b)$.

For the Fano-type bound, we again notice that
\begin{equation}
 \ent{X|Y}=\ent{E|Y}+\ent{X|E,Y}.
\end{equation}
The first term can be written as
\begin{multline}
 \ent{E|Y}=\int_\dom{Y} \binent{\perh(y)}dP_Y(y)\\=\int_{\dom{Y}_k\setminus\dom{Y}_b}\binent{\perh(y)}dP_Y(y)\\
\leq P_Y(\dom{Y}_k\setminus\dom{Y}_b)
\end{multline}
since $\perh(y)=0$ for $y\in\dom{Y}_b$ and $\perh(y)=1$ for $y\in\dom{Y}\setminus(\dom{Y}_k\cup\dom{Y}_b)$.

For the second term we can write
\begin{IEEEeqnarray}{RCL}
 \ent{X|E,Y} &=&\int_\dom{Y} \ent{X|Y=y,E=1}\perh(y)dP_Y(y)\\
&\leq& \int_{\dom{Y}_k\setminus\dom{Y}_b} \log(\overline{K}-1) \perh(y)dP_Y(y)\notag\\
&&+{} \int_{\dom{Y}\setminus(\dom{Y}_k\cup\dom{Y}_b)} \log \overline{K} dP_Y(y)
\end{IEEEeqnarray}
Now we note that
\begin{equation}
 \perh = P_Y(\dom{Y}\setminus(\dom{Y}_k\cup\dom{Y}_b)) + \int_{\dom{Y}_k\setminus\dom{Y}_b}\perh(y)dP_Y(y)
\end{equation}
which we can use above to get
\begin{IEEEeqnarray}{RCL}
 \ent{X|E,Y} &\leq& \left(\perh-P_Y(\dom{Y}\setminus(\dom{Y}_k\cup\dom{Y}_b))\right)\log(\overline{K}-1)\notag\\&&{}+P_Y(\dom{Y}\setminus(\dom{Y}_k\cup\dom{Y}_b))\log \overline{K}.
\end{IEEEeqnarray}
Rearranging and using
\begin{equation}
 P_b+P_Y(\dom{Y}\setminus(\dom{Y}_k\cup\dom{Y}_b))+P_Y(\dom{Y}_k\setminus\dom{Y}_b)=1
\end{equation}
yields
\begin{IEEEeqnarray}{RCL}
 \ent{X|Y} &\leq& 1-P_b+\perh \log(\overline{K}-1)\notag\\&&{}+P_Y(\dom{Y}\setminus(\dom{Y}_k\cup\dom{Y}_b))\left(\log\frac{\overline{K}}{\overline{K}-1}-1\right).\notag\\
\end{IEEEeqnarray}
The fact $0\leq\log\frac{\overline{K}}{\overline{K}-1}\leq1$ completes the proof.\endproof

\ifthenelse{\proofRILdimRed=1}{}{
\section{Proof of Proposition~\ref{prop:RILdimRed}}\label{app:proofRILdimRed}
}

\ifthenelse{\proofUBRelTrans=1}{}{
\section{Proof of Proposition~\ref{prop:UBRelTrans}}\label{app:proofUBRelTrans}
 The proof follows from Proposition~\ref{prop:dimTrans}, Definition~\ref{def:infodim}, and the fact that conditioning reduces entropy. We write
\begin{align}
 \relTrans{X\to Y} &= \relTrans{X\to \Y{1}, \Y{2},\dots,\Y{K}}   \\&= \frac{\infodim{\Y{1}, \Y{2},\dots,\Y{K}}}{\infodim{X}}\\
&=\limn\frac{\ent{\Yh{1}_n,\dots,\Yh{K}_n}}{\ent{\hat{X}_n}}\\&=\limn\frac{\sum_{i=1}^K \ent{\Yh{i}_n|\Yh{1}_n,\dots,\Yh{i-1}_n}}{\ent{\hat{X}_n}}\\
&\leq \limn\frac{\sum_{i=1}^K \ent{\Yh{i}_n}}{\ent{\hat{X}_n}} = \frac{\sum_{i=1}^K \infodim{\Yh{i}}}{\infodim{X}}\\&=\sum_{i=1}^K\relTrans{X\to \Y{i}}
\end{align}
where we exchanged limit and summation by the same reason as in the proof of Proposition~\ref{prop:RILDim}.
\ifthenelse{\proofUBRelTrans=1}{}{\endproof}}

\ifthenelse{\proofUBRelLoss=1}{}{
\section{Proof of Proposition~\ref{prop:UBRelLoss}}\label{app:proofUBRelLoss}
 The proof follows from the fact that $\infodim{X}=N$ and, for all $i$, $\infodim{\X{i}}=1$. We obtain from the definition of relative information loss
\begin{multline}
 \relLoss{X\to Y} = \limn \frac{\ent{\Xh{1}_n,\dots,\Xh{N}_n|Y}}{\ent{\Xh{1}_n,\dots,\Xh{N}_n}}\\
= \limn \frac{\sum_{i=1}^N\ent{\Xh{i}_n|\Xh{1}_n,\dots,\Xh{i-1}_n,Y}}{\ent{\Xh{1}_n,\dots,\Xh{N}_n}}\\
\leq \limn\frac{\sum_{i=1}^N\ent{\Xh{i}_n|Y}}{\ent{\Xh{1}_n,\dots,\Xh{N}_n}}\label{eq:proofUBRelLoss}
\end{multline}
Exchanging again limit and summation we get
\begin{equation}
 \relLoss{X\to Y} \leq \frac{\sum_{i=1}^N \infodim{\X{i}|Y}}{\infodim{X}}=\frac{\sum_{i=1}^N \infodim{\X{i}|Y}}{N}.
\end{equation}
But since $N=N\infodim{\X{i}}$ for all $i$, we can write
\begin{equation}
 \relLoss{X\to Y} \leq \frac{1}{N} \sum_{i=1}^N \frac{\infodim{\X{i}|Y}}{\infodim{\X{i}}} = \frac{1}{N} \sum_{i=1}^N\relLoss{\X{i}\to Y}.
\end{equation}
This proves the first inequality. The second is obtained by removing conditioning again in~\eqref{eq:proofUBRelLoss}, since $Y=\{\Y{1},\dots,\Y{N}\}$.
\ifthenelse{\proofUBRelLoss=1}{}{\endproof}}

\ifthenelse{\prooffanoUBRIL=1}{}{
\section{Proof of Proposition~\ref{prop:fanoUBRIL}}\label{app:prooffanoUBRIL}
Note that by the compactness of $\dom{X}$ the quantized input $\hat{X}_n$ has a finite alphabet, which allows us to employ Fano's inequality
\begin{equation}
 \ent{\hat{X}_n|Y} \leq \binent{P_{e,n}} + P_{e,n}\log\card{\partit{n}}
\end{equation}
where
\begin{equation}
 P_{e,n} = \Prob{\rec{Y}\neq\hat{X}_n}.
\end{equation}
Since Fano's inequality holds for arbitrary reconstructors, we let $\rec{\cdot}$ be the composition of the MAP reconstructor $\recmap{\cdot}$ and the quantizer introduced in Section~\ref{ssec:notation}. Consequently, $P_{e,n}$ is the probability that $\recmap{Y}$ and $X$ do not lie in the same quantization bin. Since the bin volumes reduce with increasing $n$, $P_{e,n}$ increases monotonically to $\perr$. We thus obtain with $\binent{p}\leq 1$ for $0\leq p \leq 1$
\begin{equation}
 \ent{\hat{X}_n|Y} \leq 1+ \perr\log\card{\partit{n}}.
\end{equation}
With the introduced definitions,
\begin{IEEEeqnarray}{RCL}
 \relLoss{X\to Y} &=& \limn \frac{\ent{\hat{X}_n|Y}}{\ent{\hat{X}_n}}\\
&\leq& \limn \frac{1+ \perr\log\card{\partit{n}}}{\ent{\hat{X}_n}}\\
&\stackrel{(a)}{=}&\perr\frac{\boxdim{\dom{X}}}{\infodim{X}}
\end{IEEEeqnarray}
where $(a)$ is obtained by dividing both numerator and denominator by $n$ and evaluating the limit. This completes the proof.
\ifthenelse{\prooffanoUBRIL=1}{}{\endproof}}

\ifthenelse{\proofRILmixed=1}{}{
\section{Proof of Proposition~\ref{prop:RILmixed}}\label{app:proofRILmixed}
Since the partition is finite, we can write with~\cite[Thm.~2]{Wu_Renyi}
\begin{equation}
 \infodim{X|Y=y}= \sum_{i} \infodim{X|Y=y,X\in\dom{X}_i} P_{X|Y=y}(\dom{X}_i).
\end{equation}
If $g$ is injective on $\dom{X}_i$, the intersection $\preim{y}\cap\dom{X}_i$ is a single point\footnote{We do not consider the case here that the preimage of $y$ does not intersect $\dom{X}_i$, since in this case $P_{X|Y=y}(\dom{X}_i)=0$.}, thus $\infodim{X|Y=y,X\in\dom{X}_i}=0$. Conversely, if $g$ is constant on $\dom{X}_i$, the preimage is $\dom{X}_i$ itself, so one obtains
\begin{equation}
 \infodim{X|Y=y,X\in\dom{X}_i}=\infodim{X|X\in\dom{X}_i}= \frac{P_X^{ac}(\dom{X}_i)}{P_X(\dom{X}_i)}.
\end{equation}
We thus write
\begin{IEEEeqnarray}{RCL}
 \infodim{X|Y}&=&\int_{\dom{Y}} \sum_{i} \infodim{X|Y=y,X\in\dom{X}_i} P_{X|Y=y}(\dom{X}_i)dP_Y(y)\notag\\\\
&=& \int_\dom{Y} \sum_{i: \dom{X}_i\subseteq A}\frac{P_X^{ac}(\dom{X}_i)}{P_X(\dom{X}_i)}P_{X|Y=y}(\dom{X}_i)dP_Y(y)\\
&\stackrel{(a)}{=}&  \sum_{i: \dom{X}_i\subseteq A}\frac{P_X^{ac}(\dom{X}_i)}{P_X(\dom{X}_i)}\int_\dom{Y}P_{X|Y=y}(\dom{X}_i)dP_Y(y)\\
&\stackrel{(b)}{=}& P_X^{ac}(A)
\end{IEEEeqnarray}
where in $(a)$ we exchanged summation and integration with the help of Fubini's theorem and $(b)$ is due to the fact that the sum runs over exactly the union of sets on which $g$ is constant, $A$. Proposition~\ref{prop:RILDim} completes the proof.
\ifthenelse{\proofRILmixed=1}{}{\endproof}}

\section*{Acknowledgments}
The authors thank Yihong Wu, Wharton School, University of Pennsylvania, Siu-Wai Ho, Institute for Telecommunications Research, University of South Australia, and Sebastian Tschiatschek, Signal Processing and Speech Communication Laboratory, Graz University of Technology, for fruitful discussions and suggesting material. In particular, the authors wish to thank Christian Feldbauer, formerly Signal Processing and Speech Communication Laboratory, Graz University of Technology, for his valuable input during the work on Section~\ref{sec:PBFs}.

\bibliographystyle{IEEEtran}
\bibliography{IEEEabrv,/afs/spsc.tugraz.at/project/IT4SP/1_d/Papers/InformationProcessing.bib,%
/afs/spsc.tugraz.at/project/IT4SP/1_d/Papers/ProbabilityPapers.bib,%
/afs/spsc.tugraz.at/user/bgeiger/includes/textbooks.bib,%
/afs/spsc.tugraz.at/user/bgeiger/includes/myOwn.bib,%
/afs/spsc.tugraz.at/user/bgeiger/includes/UWB.bib,%
/afs/spsc.tugraz.at/project/IT4SP/1_d/Papers/InformationWaves.bib,%
/afs/spsc.tugraz.at/project/IT4SP/1_d/Papers/ITBasics.bib,%
/afs/spsc.tugraz.at/project/IT4SP/1_d/Papers/HMMRate.bib,%
/afs/spsc.tugraz.at/project/IT4SP/1_d/Papers/ITAlgos.bib}

\end{document}